%% file: main.tex
\documentclass[aps,prb,longbibliography,amsmath,amssymb,superscriptaddress,twocolumn,10pt]{revtex4-2}
\usepackage{standalone}
\usepackage{txfonts}%
\usepackage{graphicx}%
\usepackage{dcolumn}%
\usepackage{bm}%
\usepackage{array}
\usepackage[dvipsnames]{xcolor}
\usepackage[%
    colorlinks=true,
    pdfborder={0 0 0},
    linkcolor=blue
]{hyperref}
\usepackage{chngpage}
\usepackage{appendix}
\usepackage{xspace}
\usepackage{comment}

\usepackage{tikz}
\usetikzlibrary{positioning, shapes, arrows, fadings,through}

\graphicspath{{Figures/}}

\begin{document}
\def\avs {$A$V$_3$Sb$_5$\xspace}
\def\rvs {RbV$_3$Sb$_5$\xspace}
\def\cvs {CsV$_3$Sb$_5$\xspace}
\def\kvs {KV$_3$Sb$_5$\xspace}
\def\V {$^{51}$V\xspace}
\def\Sb {$^{121}$Sb\xspace}
\def\musr {$\mu$SR\xspace}

\title{Evolving charge order in the CDW state of \avs metals}

\author{Anshu Kataria}
\affiliation{
Dipartimento di Scienze Matematiche, Fisiche e Informatiche, Universit\`a di Parma, I-43124 Parma, Italy
}

\author{Francis Pratt}
\affiliation{ISIS Pulsed Neutron and Muon Source, Rutherford Appleton Laboratory, Didcot OX11 0QX, U.K.}

\author{Peter J. Baker}
\affiliation{ISIS Pulsed Neutron and Muon Source, Rutherford Appleton Laboratory, Didcot OX11 0QX, U.K.}

\author{Stephen Cottrell}
\affiliation{ISIS Pulsed Neutron and Muon Source, Rutherford Appleton Laboratory, Didcot OX11 0QX, U.K.} 

\author{Miki Bonacci}
\affiliation{PSI Center for Scientific Computing, Theory and Data, 5232 Villigen PSI, Switzerland}

\author{Andrea Capa Salinas} 
\affiliation{Materials Department, University of California Santa Barbara, Santa Barbara, California 93106, USA}

\author{Stephen D. Wilson}
\affiliation{Materials Department, University of California Santa Barbara, Santa Barbara, California 93106, USA}

\author{Zurab Guguchia}
\affiliation{PSI Center for Neutron and Muon Sciences CNM, 5232 Villigen PSI, Switzerland}

\author{Samuele Sanna}
\email[Contact author: ]{s.sanna@unibo.it}
 \affiliation{
Dipartimento di Fisica e Astronomia  ``A. Righi'', Universit\`a di Bologna, I-40127 Bologna, Italy }

\author{\hspace{1mm}Pietro Bonf\`a}
\email[Contact author: ]{pietro.bonfa@posteo.net}
\affiliation{Dipartimento di Fisica, Informatica e Matematica, Universit\`a di Modena e Reggio Emilia, Via Campi 213/a, 41125 Modena, Italy}
\affiliation{CNR-NANO S3—Istituto Nanoscienze, I-41125 Modena, Italy }

\date{\today}

\begin{abstract}

\avs kagome metals are characterized by intertwined electronic and structural orders, which motivated extensive studies in recent years. Yet the details of the electronic state preceding the superconducting phase remain poorly understood.
Here we extend our previous investigation [Phys. Rev. Research 7, L032046 (2025)] of \rvs using avoided level crossing (ALC) muon-spin spectroscopy to the $A$ = Cs and K systems. Consistent with our previous study, we identify a second transition whose origin cannot be attributed solely to an internal magnetic field, indicating the involvement of an additional electronic mechanism that subtly modifies the charge distribution within the V plane.
In particular, the ALC results point towards an additional charge modulation taking place within the charge density wave (CDW) phase and occurring at $T^{*}<T_{CDW}$ for $A$ = Cs and Rb, or in the vicinity of $T_{CDW}$ for $A$ = K.

\end{abstract}
\maketitle

\section{Introduction}

Vanadium-based kagome metals \avs ($A$ = K, Rb and Cs) have attracted significant research interest because of the intricate nature of their electronic states, as testified by their rich phase diagram \cite{kagome.first,wilson2024v3sb5,zhao2021cascade,1g9n-wm38}. This complexity originates from the interplay of electronic correlations, non-trivial band topology and Van Hove singularities near the Fermi level, giving rise to a variety of phase transitions, with the most accurately characterized being the high-temperature charge density wave (CDW) ($T_{CDW}\sim$ 94~K for \cvs, $T_{CDW} \sim$ 103~K for \rvs, $T_{CDW}\sim$ 78~K for \kvs) and superconductivity, at around 2~K. Other symmetry breaking electronic transitions 
\cite{ortiz2020cs,xu2022three, denner2021analysis, doi:10.1126/sciadv.abb6003, yu2021concurrence, PhysRevB.111.L041109, kang2022twofold,mielke2022time,khasanov2022time,trsb_optical,chen2021roton, li2022rotation,jiang2021unconventional,cheng2025broken,PhysRevB.105.195136, zhao2021cascade,luo2022possible} 
have been observed at or below the CDW order,
but their nature and origin 
is still debated.

From the structural point of view, despite sharing the same hexagonal structure at high temperature, \avs exhibit different lattice and charge modulations within the CDW phase. 
While a common motif, also supported by recent theoretical investigations, is the formation of the staggered tri-hexagonal modulation \cite{PhysRevResearch.5.L012017, 4r8x-j3nd, park2023condensation, PhysRevB.107.184106}, for the Cs system, a competition of 2$c$ and 4$c$ modulation along the $c$-axis, possibly involving also star-of-David distortions, is suggested \cite{PhysRevX.11.041030, PhysRevB.105.155106,PhysRevB.105.195136, PhysRevMaterials.8.093601}. A unidirectional charge modulation with periodicity 4$a_0$, breaking the six fold kagome plane symmetry, has further been observed in Cs and Rb systems, while absent in the K system \cite{ zhao2021cascade,li2022rotation,luo2022possible,PhysRevMaterials.5.L111801,PhysRevX.13.031030,trsb_optical,li2023unidirectional,PhysRevB.104.035131,trsb_optical}. Additionally, \cvs also exhibits an electric magneto-chiral anisotropy which seems to be negligible for \kvs, irrespective of their similar electronic structure \cite{guo2022switchable,guo2024distinct}. 
These alkali-metal-dependent properties suggest a non-trivial role of the spacer layer in determining the ground state of these materials, thereby affecting their electronic response.

Even more intriguing is the possibility of time-reversal symmetry (TRS) breaking at or below the CDW transition.
Early studies, including muon spin rotation and relaxation (\musr) measurements reporting an increased spin relaxation rate 
\cite{mielke2022time,trsbkagome,Kenney_2021} and its field and muon penetration depth enhancement \cite{PhysRevResearch.4.023244,graham2024depth}, magnetic-field-induced switching of the CDW Bragg peaks intensity in scanning tunneling microscopy (STM) \cite{jiang2021unconventional,trsb_optical}, anisotropic and non-reciprocal electronic magneto-transport \cite{doi:10.1126/sciadv.abb6003,wei2024three}, circular dichroism \cite{cha2026evidence} and observation of a magnetic peak in polarized neutron diffraction measurements \cite{PhysRevB.110.195109}, have suggested the presence of a non-trivial magnetic response that may be compatible with the presence of very small magnetic moments under zero field conditions. However, this interpretation has been questioned due to the absence of Kerr rotation
\cite{PhysRevLett.131.016901,farhang2023unconventional,PhysRevMaterials.8.014202, PhysRevB.105.045102}.
Nevertheless, a null Kerr response does not exclude TRS breaking for compensated or staggered magnetic orders. Moreover, recent theoretical work showed that Kerr and anomalous Hall responses depend sensitively on the orbital-flux pattern, and the absence of Kerr effect cannot exclude the broken TRS \cite{PhysRevB.111.L041109}.

A recent theoretical study based on vestigial order suggested that the symmetry-breaking phase around $T^*\lesssim T_{CDW}$ may originate from subtle perturbations by the probe and a strong sensitivity to external conditions like magnetic field \cite{ingham2025vestigial}. While this theory successfully explains several experimental observations \cite{huang2026magnetic,trsb_optical,guo2022switchable,PhysRevX.14.031015,le2024superconducting,guo2024correlated} including the absence of Kerr rotation \cite{PhysRevLett.131.016901,farhang2023unconventional,PhysRevMaterials.8.014202}, it also raises questions about the intrinsic nature of the electronic state in the intermediate temperature regime, thus motivating a detailed investigation under zero and near-zero-field conditions.

In this work, we employ avoided level crossing (ALC) \musr measurements performed under weak longitudinal field (LF) conditions, which probe the splitting of nuclear levels induced by the electric field gradient through the dipolar coupling between muons and nuclei \cite{le2011muon}. A so-called ``resonance'' happens %
when the nuclear splitting matches the Zeeman splitting of the muon spin energy levels, leading to a cross relaxation process \cite{Cox, Kreitzman1986}. The technique is sensitive to the nuclei surrounding the implanted muon and can selectively probe different nuclei by tuning the muon's Zeeman splitting with the applied magnetic field. %
This approach proved to be effective in disentangling charge and spin degrees of freedom for $A$=Rb in Ref.~\cite{bonfa2024unveiling} and here we extend the same investigation to the $A$=Cs and K systems.

In addition, we performed high-statistics zero-field (ZF)-$\mu$SR measurements on the \kvs system, confirming the previous observation \cite{Kenney_2021, mielke2022time} of an increase in the muon relaxation rate at low temperatures. 
Surprisingly, a similar temperature dependence is also observed in the parameters describing the ALC resonances across all \avs.
Combining experimental measurements with first-principle calculations, we present a comprehensive study of \avs below $T_{CDW}$ demonstrating that the ALC-\musr transition cannot be accounted for solely by the emergence of internal (hyperfine) fields, but instead points to an additional electronic transition and underscores the rich landscape of intertwined electronic and magnetic phenomena in kagome superconductors

\section{Results}\label{sec:results}

\subsection{ZF-$\mu$SR}
ZF-$\mu$SR measurements were performed on high-quality \kvs powdered samples \cite{kagome.first} at the ISIS facility using the EMU spectrometer.
ZF-$\mu$SR data for \cvs have already been acquired at the same facility by Z. Shan $et. al.$ and are presented in Ref. \cite{cvs_zfmuon}. For the sake of consistency, we re-analyzed their data \cite{zf_muon_repository} with the model described below and obtained nearly identical results \footnote{ We note that the abstract of Ref.~\cite{zf_muon_repository} reports that local fields are found to be ``dynamic'' in nature. This conclusion is a consequence of the unlucky choice of applying a 5~mT LF field done by authors, who, by chance, chose the applied field precisely where the dip of the ALC resonance is. This led them to believe that the persistent relaxation was due to dynamic magnetism after comparing with a single other measurement at 50 mT. Here we show that this is not the case thank to the finely spaced LF measurements used to perform the ALC experiment.}.

\begin{figure}
\includegraphics[width=0.5\textwidth]{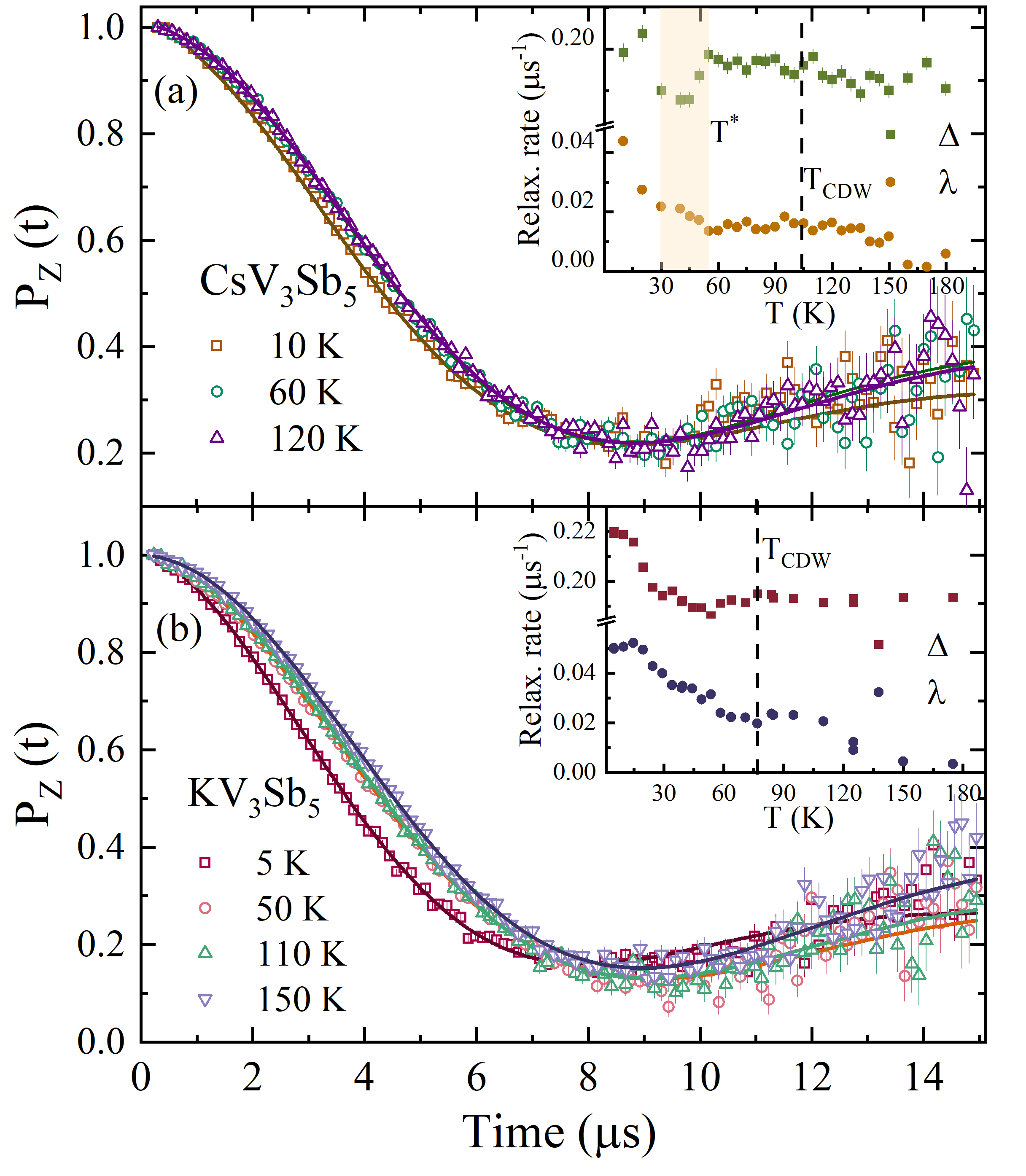}
\caption{ZF muon asymmetry, normalized to one, for (a) \cvs and (b) \kvs at different temperatures. Data in panel (a) are from \cite{cvs_zfmuon}. The solid lines are best fit to the Eq.~\ref{eq:zffit}. The insets show the relaxation rates, $\Delta$ and $\lambda$, as a function of the temperature. The dashed vertical lines in the insets represent the CDW transitions and the shaded region for \cvs represents the temperature range over which the relaxation varies and anomalies are observed in ALC measurements.}
\label{fig:zf_both}
\end{figure}

Time dependent asymmetry at selected temperatures are shown in Fig.~\ref{fig:zf_both} for both Cs and K compounds.
A clear difference between the high (150~K) and low temperature (5~K) asymmetries is observed for \kvs, while the variation of the depolarization as a function of temperature is much smaller in \cvs. To track the temperature dependence of the muon spin relaxation rate, we fitted the ZF asymmetries to the following phenomenological model
\begin{equation}
    A(t) = A_0 \Bigg[ \frac{1}{3} + \frac{2}{3} \left( 1 - {\Delta}^2 {t}^2 \right) e^{-\frac{1}{2}\Delta^2 t^2} \Bigg]e^{-\lambda t} + A_{b} \label{eq:zffit}
\end{equation}
where the first term is the Kubo-Toyabe function with $\Delta$ representing the width of the local field distribution at the muon site, $\lambda$ is an additional Lorentzian relaxation rate and $A_0$ and $A_{b}$ are the coefficients, identifying a relaxing and baseline component, respectively.  The insets of  Fig.~\ref{fig:zf_both} depict the temperature-variation of two relaxation rates, $\Delta$ and $\lambda$. 

In \cvs, $\lambda$ deviates from zero at about 150~K and it further suddenly increases at around 45 K ($T^*$). This second anomaly is accompanied by a dip-like feature in the $\Delta$ parameter. This behavior is similar to the one previously observed in \rvs \cite{bonfa2024unveiling} and might be due to a small correlation between the two fitting parameters.

\kvs shows a more pronounced deviation between the low-T and high-T relaxation rates and a similar two-step trend.
Nonetheless, at odds with \cvs, the anomaly in the Lorentzian contribution ($\lambda$) appears in the vicinity of $T_{CDW}$ transition, followed by a continuous increase down to 15~K, where it eventually saturates. On the contrary, the $\Delta$ parameter increases only below 30~K.

These behaviors are in agreement with the previous ZF-$ \mu$SR measurements \cite{Kenney_2021, mielke2022time}, but our new data provide an extended time window that allows a fine characterization of the transitions observed in ZF-\musr.

The increase of both relaxation rates for both compounds is always smaller than 0.03~$\mu\text{s}^{-1}$.
Assuming that this variation originates from an electronic contribution that manifests as a local field at the muon site through hyperfine interactions and adds to the contribution coming from nuclear moments, the total local field distribution at the muon site would increase by $\approx$ 0.04 mT \footnote{This value is obtained as either $\lambda/\gamma_\mu$
which is the half width at half maximum of a static Lorentzian field distribution or as $\Delta/\gamma_\mu$ for a Gaussian distribution}. 

\subsection{Avoided Level Crossing}

\begin{figure}
\includegraphics[width=0.45\textwidth]{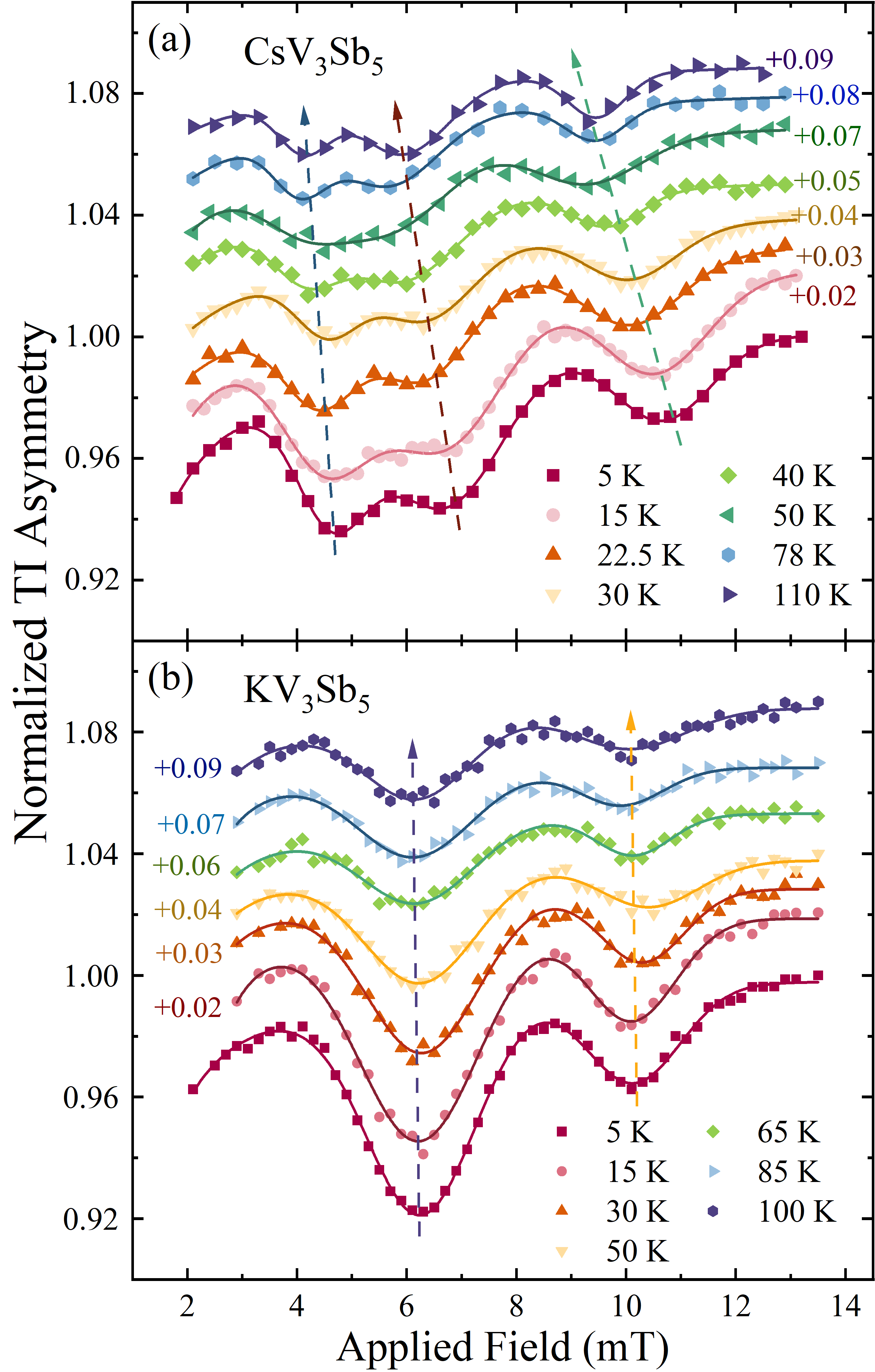}
\caption{ Normalized TI asymmetry variation with applied field at different temperatures up to above the $T_{CDW}$ for (a) \cvs and (b) \kvs. Temperature-dependent resonance spectra are vertically shifted by the amounts reported in the picture for clarity. The solid colored lines represent the best fit using Eq.~\ref{eq:alcfit}. The dashed arrow lines are guides to the eye to showcase the variation of resonance fields with increasing temperature.}
\label{TIasyms_cvs}
\end{figure}

ALC measurements were also performed on the same spectrometer with LF ranging from 2.1 and 13.3~mT.
These measurements can be analyzed with two, equivalent, methods.
The time-dependent asymmetry can be fit to a stretched exponential function \cite{Kreitzman1986}, which can effectively capture the variation of the relaxation rate in the proximity of the ALC resonance.
Alternatively, the muon asymmetry can be integrated over the entire time window used for the data acquisition. At resonance, the faster muon depolarization leads to a reduction in the time-integrated (TI) asymmetry, producing a dip feature in the spectrum. This latter approach is discussed in the main text while the former is reported in the supplementary material (SM) \footnote{See Supplemental Material for the details on the \musr{} experimental method, the ALC relaxation analysis, first principles and spin Hamiltonian simulations, muon site information, simulation of ZF polarization functions and comparison with experimental data, accuracy of the minimal model.
}

Fig. \ref{TIasyms_cvs} shows the ALC-\musr results in terms of TI asymmetry presented as a function of applied LF at different temperatures. A shift in the resonance field with temperature, indicated by the dashed arrows, is clearly seen for \cvs{} (Fig. \ref{TIasyms_cvs}(a)), whereas only a minimal change is noted for the K system (Fig. \ref{TIasyms_cvs}(b)). At the lowest temperature, 5~K, \cvs exhibit two overlapping resonance dips at 4.63(3) and 6.69(4) mT, which are followed by another resonance at a higher field, 10.66(3) mT.
For \kvs, only two sharp resonance dips are present at 6.22(2) and 10.10(3) mT. 
At the same time, a significant increase in the area of the resonances is instead observed for both the compounds.
Notably, the recovery of the polarization in LF measurements above the resonance field saturates at approximately the same applied field for both systems at all temperatures. This observation, together with the excellent agreement between the experiment data and the prediction obtained entirely from nuclear magnetism (described in the next section), rules out the presence of dynamical magnetism of electronic origin within the \musr{} time window over entire temperature range explored for both \cvs and \kvs.

To analyze the experimental results, the field-dependent TI asymmetry has been fitted by considering a Gaussian distribution for each resonance and a power-law behavior for the background, expressed as \cite{bonfa2024unveiling};
\begin{equation}
    G(H) = G_{0} \left(1-\frac{\tau}{\mu_0 H^{n}}\right) - \sum_{i=1}^{m} 
    \frac {A_{i}}{\sigma_{i} {\sqrt {2\pi }}}\exp \left(-{\frac {1}{2}}\frac {(\mu_0H-B^{\mathrm{res}}_{i} )^{2}}{\sigma_{i} ^{2}}\right) \label{eq:alcfit}
\end{equation}
where $\mu_0H$ is the applied field and $G_{0}$ corresponds to the integrated asymmetry as the field goes to infinity. The first term in the above equation accounts for the muon-nuclei dipolar interaction without any resonance and it is purely based on a phenomenological approach, where $\tau$ and $n$ are the background fitting parameters. The parameter $m$ accounts for the number of resonance dips observed, which is 3 for Cs and 2 for K system. Each Gaussian function is defined by three parameters, $A_i$, $B_i^{res}$ and $\sigma_{i}$, corresponding to the area, field and width of the resonances, respectively.

\begin{figure*}
\includegraphics[width=1.0\textwidth]{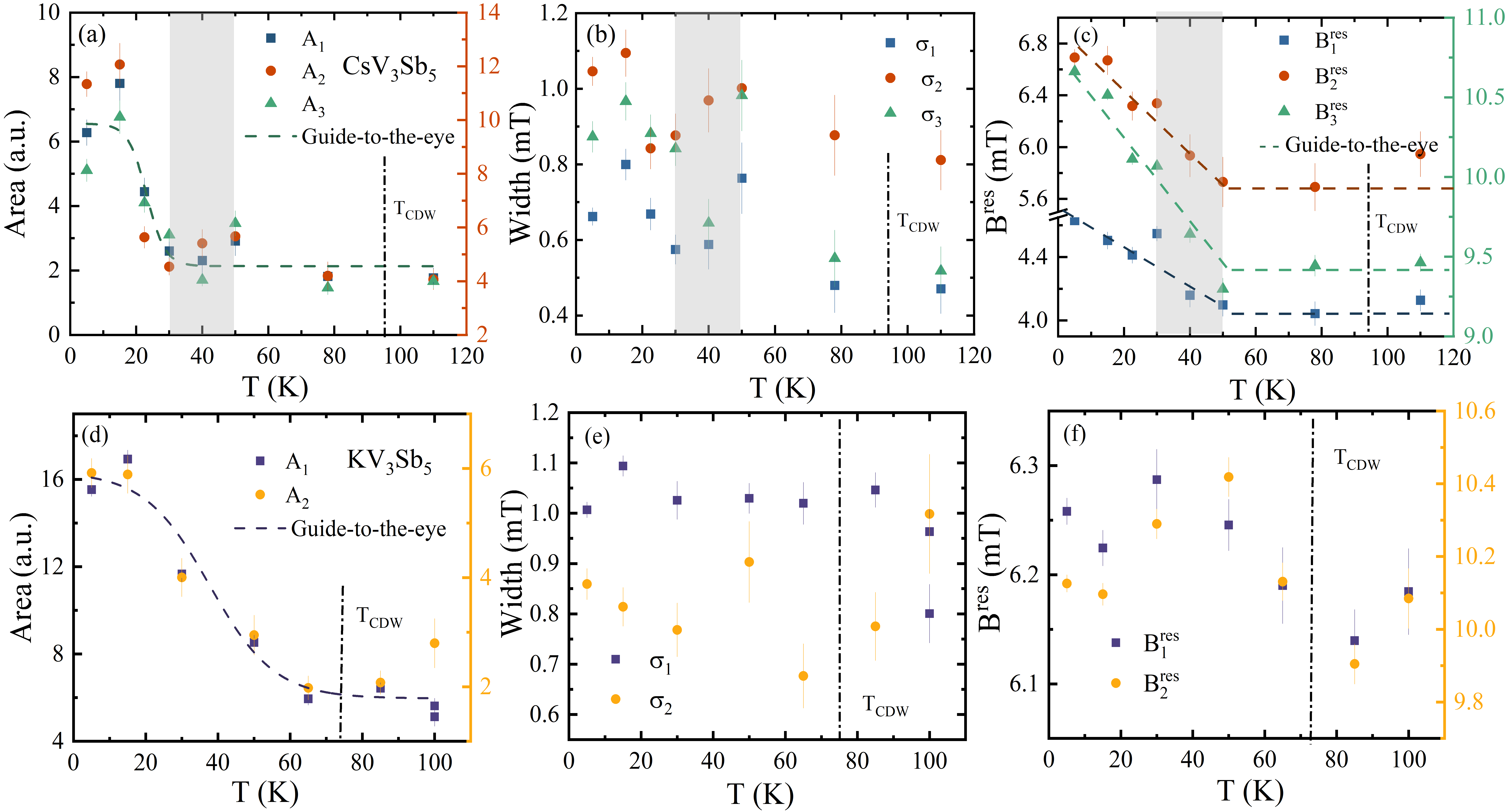}
\caption{The top panel shows the temperature dependence of the fitting parameter for the \cvs system, (a) area (b) width and (c) resonant field of three resonances. The bottom panel presents the same fitting parameters for \kvs system. The dashed lines, color coded according to the respective parameters, serve as guide-to-the-eye for visualizing their temperature-dependent evolution. The vertical black dot-dashed line indicates the CDW transition temperature, $T_{CDW}$. Finally, the shaded region in panels (a-c)  corresponds to the temperature range where anomalies are observed in both ZF and ALC $\mu$SR measurement. %
}
\label{fig:fitparam}
\end{figure*}

The temperature variation of the fitted parameters for both systems are shown in Fig. \ref{fig:fitparam}. For \cvs, the resonance area exhibits a minimal change around the CDW transition (marked by dot-dashed vertical line,  Fig. \ref{fig:fitparam}(a)), while a significant increase is observed at a much lower temperature, $T^*_1\approx$ 30 K, followed by a saturation at low-temperature. 
The resonances width, $\sigma_i$, is almost temperature independent (Fig. \ref{fig:fitparam}(b)), while a seemingly linear increment with decreasing temperature is noted for the resonance field, $B_i^{res}$, of all three dips (Fig. \ref{fig:fitparam}(c)). The corresponding increase, however, occurs at a slight higher temperature,  $T^*_2\approx$ 50 K. 
The two temperatures are represented by a solid grey band, which is also reported in the inset of Fig.~\ref{fig:zf_both}(a) to simplify the comparison with ZF results, where two anomalies in the $\lambda$ parameter are observed at the same temperatures.

Fig.~\ref{fig:fitparam}(d-f) depict the temperature dependence of the same fitting parameters for \kvs.
A systematic increase in the area of the resonances is observed close to $T_{CDW}$, at $T^* \approx$ 60 K, with a saturation at low-temperature (Fig.~\ref{fig:fitparam}(d)). 
As noted earlier for \cvs, also in this case $T^*$ roughly corresponds to the temperature where the ZF relaxation rate $\lambda$ increases.
In contrast to the $A$=Cs and Rb cases \cite{bonfa2024unveiling}, the resonance field and its width do not show a clear temperature dependence (Fig. \ref{fig:fitparam}(e-f)).

It is finally noted that the estimated increase in the total field distribution at the muon site obtained from ZF measurements is negligible when compared to the 1~mT shift shown in Fig.~\ref{fig:fitparam}(c) for \cvs, while it may still be compatible with the absence of a detectable shift in \kvs (Fig.~\ref{fig:fitparam}(f)). Yet, as detailed below, this would not explain the large increase in the  area of the resonances observed for both systems.
Notably, \rvs also shows the same systematic increase in the resonance area, along with the resonance field, upon decreasing temperature (Fig. S3 of SM). 

Summarizing, an increasing resonance dip is consistently observed across the entire \avs family, although with different onset temperatures, while for A=Cs (and Rb), the resonant field is also varying across $T^*$.

\subsection{First principles simulations}

To obtain a microscopic understanding of the \musr results, we employed first-principle density functional theory (DFT+$\mu$ \cite{blundell2023dft+}) simulations 
to identify muon sites and to parametrize the muon-nuclei spin Hamiltonian. We follow the same approach used for \rvs in Refs.~\cite{bonfa2024unveiling,graham2024depth}, with further computational details provided in the SM. 

For both systems, we considered the $\pi$-shifted tri-hexagonal phase, which is reported to be the most stable structure~\cite{4r8x-j3nd} below CDW transition for both $A$=Cs and K.
The same three lowest energy and symmetrically inequivalent muon sites already reported for \rvs are also found for the other two compounds of this family.
All three muon sites have similar atomic coordination and energy differences smaller than 0.2~eV (further details are provided in the SM).
In light of the very small geometric differences among them and the minor variation of the electric field gradients (EFG) at the neighboring nuclear sites, we only consider one representative for the entire set and refer to it as \textit{site~1} in the following analysis.

\begin{figure*}
\includegraphics[width=1.0\textwidth]{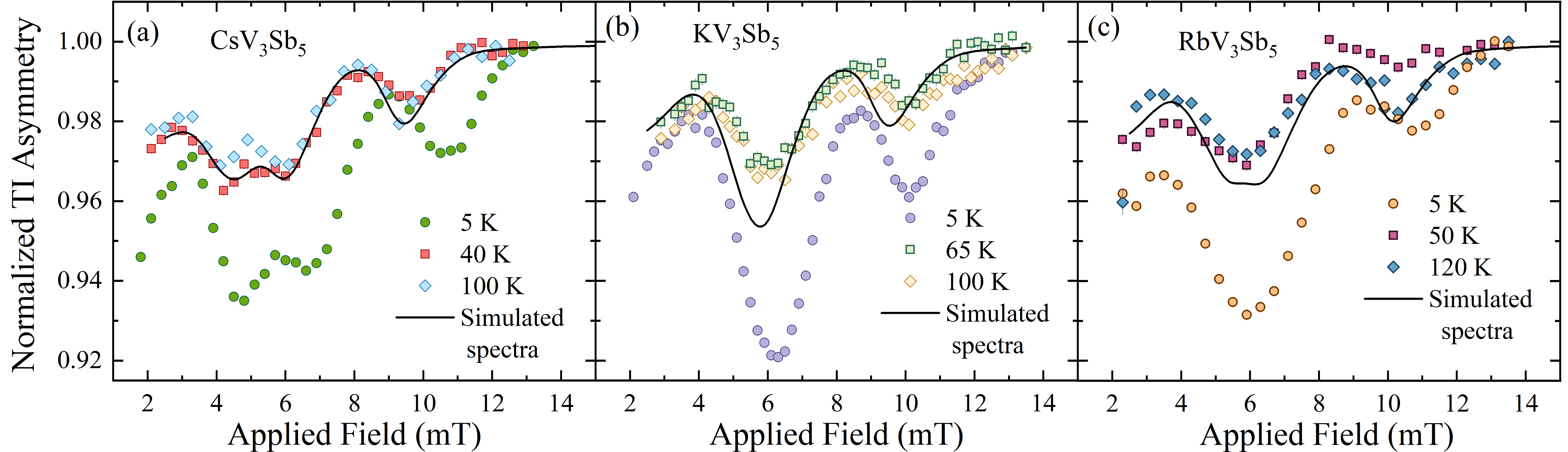}
\caption{Experimental and simulated ALC-$\mu$SR spectra, considering the cluster of the nearest Sb atom and other six nearest V atoms (see SM) roughly forming a hexagon around the muon stopping site for (a) \cvs, (b) \kvs and (c) \rvs (data from \cite{bonfa2024unveiling}). The y-axis represents the (normalized) time-integrated asymmetry.}
\label{alc_simulation}
\end{figure*}

From the inspection of the (muon perturbed) EFG tensors at the nuclear sites close to the muon, we conclude that the quadrupolar splitting of V nuclei generates the resonances in the magnetic-field range 2-13~mT probed experimentally \footnote{More details in the SM. The muon-induced perturbation of the EFG at next-nearest Cs atom is sizable ($\lesssim 40$~\%, while the one for the V atoms is much smaller ($<10$~\%)). Interestingly the EFGs at the nuclei far from the muon in the supercell calculations are in good agreement with the experimental results of NQR measurements \cite{PhysRevB.107.184106,luo2022possible}.}.
The expected resonance fields from other quadrupolar nuclei (Sb and Cs/K) are either at much higher or lower values, respectively, consistent with previous findings for $A$=Rb \cite{graham2024depth,bonfa2024unveiling}.

Having characterized both the geometry and the EFGs of the nuclei surrounding the implanted muon at site 1, we estimated the ZF and LF muon polarization functions for the two systems by computing the time evolution of muon spin using the Celio approach \cite{Celio_1986, undi} and considering the following muon-nuclei spin Hamiltonian:
\begin{equation}
    \mathcal{H} = \mathcal{H}_{Z,\mu} + \sum_{i}^{N} \left[  \mathcal{H}_{Z,i} + \mathcal{H}_{D,i} +  \mathcal{H}_{Q,i} \right].
    \label{eq:spinh}
\end{equation}
In Eq.~\ref{eq:spinh}, $\mathcal{H}_{Z,\mu} = -\hbar \gamma_{\mu} \mathbf{I}_{\mu}\cdot (\mathbf{B}_{\mathrm{ext}} + \mathbf{B}_{\mathrm{int}})$ is the Zeeman term for the muon with spin $I_\mu = 1/2$ and gyromagnetic ratio $\gamma_\mu$, possibly subject to the external field $\mathbf{B}_{ext}$ or the internal field $\mathbf{B}_{\mathrm{int}}$. $\mathcal{H}_{Z,i} = -\hbar \gamma_{i} \mathbf{I}_{i}\cdot (\mathbf{B}_{\mathrm{ext}} + \mathbf{B}_{\mathrm{int}})$ is the same for nucleus $i$ with spin $I_i$ and gyromagnetic ratio $\gamma_i$ and $N$ neighboring nuclei are considered. The dipolar interaction between the muon and the nuclei is given by
\begin{equation}
\mathcal{H}_{D,i} = \frac{\mu_0 \hbar^2 }{4\pi}\gamma_{i}\gamma_{\mu} \left( \frac{\mathbf{I}_{i}\cdot\mathbf{I}_{\mu}}{r^3} -  \frac{3(\mathbf{I}_{i}\cdot\mathbf{r})(\mathbf{I}_{\mu}\cdot\mathbf{r})}{r^5} \right),     
\end{equation}
with $r$ being the distance between the $i$-th nucleus and the muon, while the quadrupolar interaction for nucleus $i$ reads as 
\begin{equation}
\mathcal{H}_{Q,i} = \frac{eQ_{i}}{6I_{i}(2I_{i}-1)} \sum_{{\alpha},{\beta}{\in}\{{x,y,z}\}} V_{i}^{{\alpha}{\beta}} \left[ \frac{3}{2}\left( I_{i}^{\alpha}I_{i}^{\beta} - I_{i}^{\beta} I_{i}^{\alpha}  \right) - \delta_{{\alpha}{\beta}}I_{i}^{2} \right]
\end{equation}
where $Q_i$ is electric quadrupolar moment and $V_i ^{\alpha \beta}$ is EFG tensor (obtained from first principles and including the perturbation induced by the muon) at the $i$-th nucleus.

The simulated ZF polarization functions 
lie exactly in between the (very similar) low and high-temperature depolarization curves for \cvs, whereas, for \kvs, it follows the trend measured above 60~K very closely (complete details are provided in the SM). 

The same method is also used to simulate the time-integrated asymmetries at the different fields from which ALC-$\mu$SR spectra are constructed. 
Experimentally observed resonances above and below $T_{CDW}$ and $T^*$ for all three \avs systems are shown in Fig. \ref{alc_simulation} together with the first principles predictions.

A very good quantitative agreement with the simulated spectra is observed for the data above $T^*$ inside the CDW state (40 K for \cvs and \rvs, and 65~K for \kvs)  \footnote{We stress that both these results and the ones reported in the SM for comparing ZF measurements with the computational prediction are completely parameter free: the only data processing is the removal of a small background component from ZF asymmetries and the normalization of both ZF and ALC results based on the total asymmetry observed experimentally. On a side note, we mention that we intentionally consider TI asymmetry instead of the cleaner results that can be obtained with the field-derivative procedure described in Ref.~\cite{bonfa2024unveiling} in order to allow absolute quantitative comparison}.
At base temperature, on the other hand, the area of the resonances is much larger than the \textit{ab initio} prediction.
In a static picture, the area of the ALC dip is governed by the dipolar interaction between the muon and V nuclei. For a muon at site 1, the observed evolution cannot be due to anharmonic corrections, since they lead to variations of the muon position smaller than $10^{-3}$~\AA{} in the temperature interval of interest.
A gradual shift of the muon's equilibrium position along the $c$-axis can also be excluded, as this would substantially change the relaxation rate of ZF asymmetries (see section VI in SM). This suggests that a more subtle mechanism must be at play, and, while its full explanation is beyond the aim of the present work, we show that the observed effect of enhanced resonance area ($i.e.,$ increased relaxation rate) cannot be of magnetic origin in the next section.

\section{Discussion}

\begin{figure}
    \centering
    \includegraphics[width=0.99\linewidth]{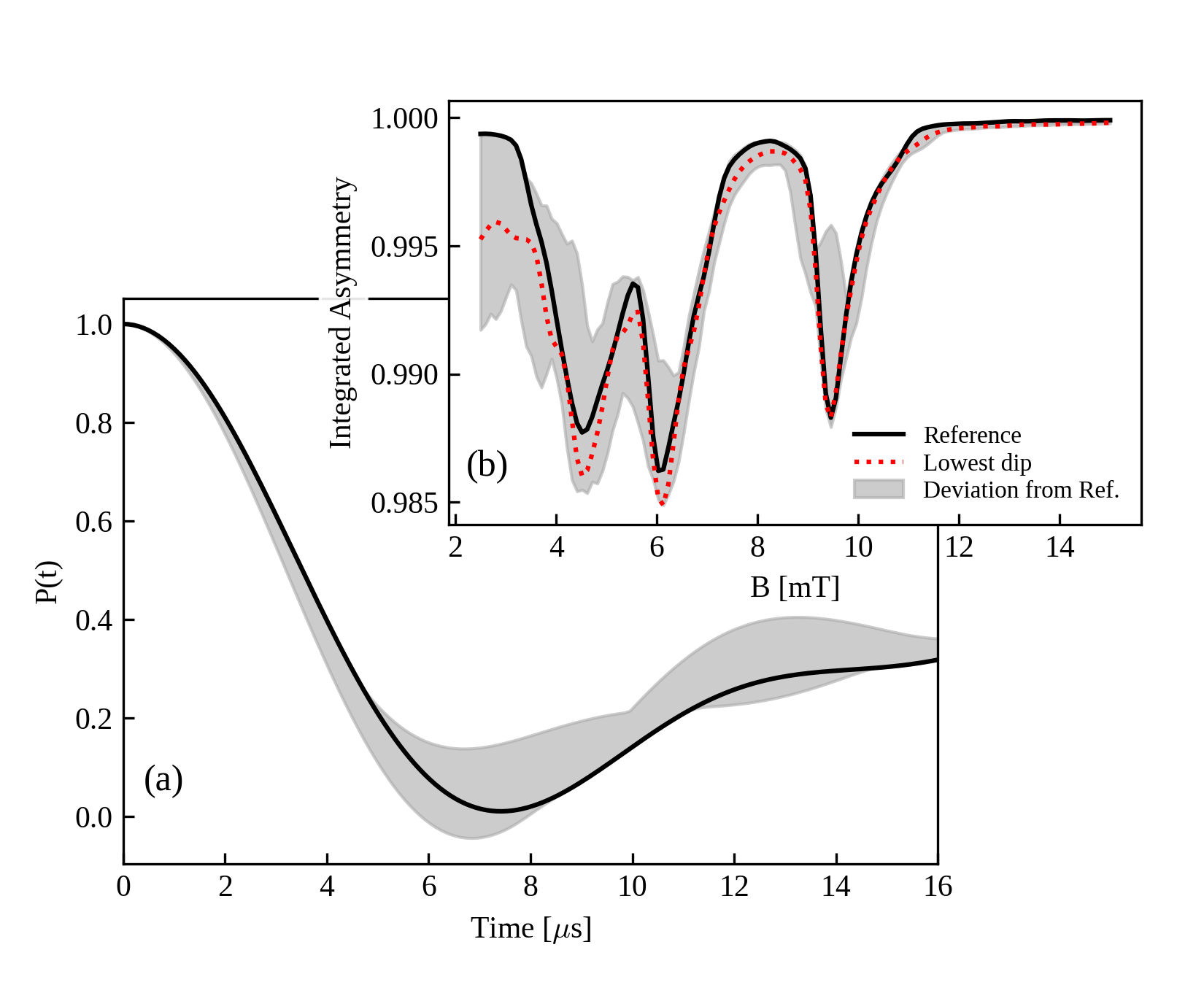}
    \caption{Perturbation induced by local fields at the muon and V sites on ZF and ALC measurements. The black line shows the results obtained with the minimal model described in the text. The shaded area highlight the changes induced by the presence of local (hyperfine) fields of electronic origin at the muon and the V nuclei. The dotted red line in panel (b) shows the one simulation with the lowest dip resulting from the presence of local fields.}
    \label{fig:numtest}
\end{figure}

\begin{figure*}[t]
  \input{tikzfig}
  \caption{Graphical representation of phase transitions described in this work. The grey shaded area identifies the SC and CDW transitions, while the orange bar shows the temperature interval where ALC-\musr shows electronic transitions. The numbers in square brackets are references to manuscripts reporting anomalies or transitions at temperatures below $T_{CDW}$ using techniques other than \musr, namely nuclear magnetic resonance~\cite{nie2022charge, Song2022, feng2026nmrevidenceloopcurrentstate}, Raman spectroscopy~\cite{PhysRevB.105.155106}
  , scanning tunneling microscopy~\cite{acs.nanolett.4c01050, li2023unidirectional,zhao2021cascade}, tuning fork resonator \cite{Gui2025}, reflective studies \cite{PhysRevB.105.245123} and (various) transport measurements \cite{Wang_2023, PhysRevX.14.031015, PhysRevB.105.L201109,doi:10.1126/sciadv.abb6003,jiang2023observation,guo2022switchable}. Notice that reports of additional transitions occurring in conjunction or above the CDW are intentionally omitted.}\label{fig:summary}
\end{figure*}
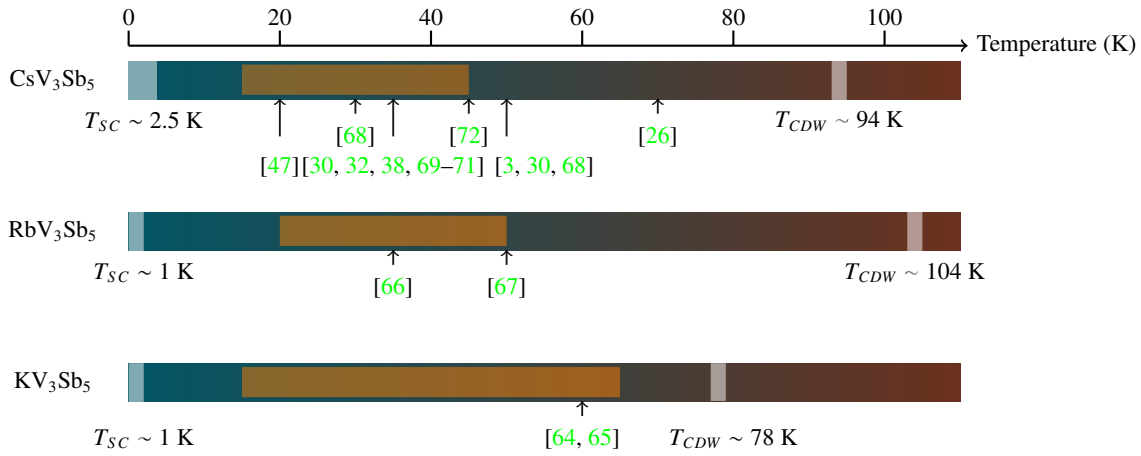

A clear shift in the resonance field with temperature is observed in the ALC-$\mu$SR results for \cvs (Fig.~\ref{fig:fitparam}(c)),  similar to what has already been reported for \rvs \cite{bonfa2024unveiling}. The conclusion of our previous work \cite{bonfa2024unveiling}, $i.e.,$ that the variation of the resonance field in ALC measurements can only be attributed to a rearrangement of the charge distribution on V nuclei, also applies for \cvs{}.
On the other hand, for \kvs, a shift in resonance field is barely observed in Fig.~\ref{fig:fitparam}(f), making the analysis less straightforward.

In all three \avs systems, however, a substantial increase in the area of the resonance dip is observed for $T^* \lesssim T_{CDW}$. 
A similar trend is also extracted from the analysis of the relaxation rate of our longitudinal field measurements (see SM). As already mentioned, this approach yields physically equivalent information as the time-integrated area, and indeed shows the same unexpected increase of a factor 2 to 3 with decreasing temperature.

To clarify this point, we start from the quantitative agreement between computational and experimental results obtained for the data collected above $T^*$, in both ZF and ALC experiments (see Fig.~\ref{alc_simulation} and Fig.~S5).
These predictions contain contributions from multiple transitions of each of the 6 V nuclear spins present in the Hilbert space, making their direct interpretation rather complicated.
In our previous work, we showed that the resonance can be approximated reasonably well by considering just a single V nucleus (together with an appropriate scaling of a factor 6 to account for the remaining 5 V neighbors). 
This approximation also holds for the present investigation as described in detail in the SM. At the same time, the ZF relaxation rate is instead primarily governed by the nearest neighboring nuclei, which are Sb and the alkali A isotopes here. %
Therefore, a minimal model based on Eq.~\ref{eq:spinh}, capable of describing both ZF and ALC measurements, requires the inclusion of the nearest neighboring Sb and $A$ nuclei and one out of six V nuclei. We use this approach to verify that the possible appearance of static internal fields \footnote{This enters as $\mathbf{B}_{int}$ in the terms of Eq.~\ref{eq:spinh} and must be static as our LF measurements rule out the presence of dynamical contributions on the \musr timescale} of electronic origin coupled through hyperfine interactions with V nuclei and the muon is not sufficient to yield the observed increase in the resonances dip in ALC measurements. 
This is naturally expected, since the additional hyperfine field can only change the splitting of the nuclear and muon levels ($i.e.,$ the resonance field) and not the strength of so-called flip-flop terms that govern cross-relaxation between the muon spin and surrounding nuclei. Unfortunately, analytical models are limited to simple configurations, so we resort to numerical methods.

We constrain the absolute values of local fields at the muon site with the value obtained from ZF measurements and that at the V nuclei with NMR results.
We finally consider arbitrary directions and uniformly distributed magnitudes of the local fields within the given constraints.
In practice, for the muon, the maximum local field magnitude is set to 0.2~mT, a value substantially larger than the one obtained from the increase of the local field distribution at muon site observed below $T^*$ in ZF and presented in section \ref{sec:results}-A. This is selected in order to provide a safe upper bound.
For the local field at V nuclei, we set the maximum value at 3~mT, since this would give rise to a shift larger than 30~kHz, well within the accuracy of the results already discussed in literature \cite{Song2022,PhysRevResearch.5.L012017,luo2022possible, 91c9-z267}.

The outcome of the numerical analysis is shown in Fig.~\ref{fig:numtest}, where 3000 simulations of random local fields have been performed using the minimal model for \cvs{}  \footnote{The alkali atom has a very minor role in the analysis of the ALC results based on the minimal model: the equilibrium distances between the muon and the neighboring atoms slightly change (see SM) as a function of $A=$K,Cs, and the nuclear moment of K or Cs only contributes to the zero field relaxation rate.}.
The black lines in Fig.~\ref{fig:numtest} show the reference with local fields set to zero. The shaded area highlights the maximum departure from the reference. 
As shown in Fig.~\ref{fig:numtest}a, in ZF the relaxation rate increases at early times, while a more complex behavior is observed beyond 7~$\mu$s, which is however, of limited interest since it is directly affected by the choice of including only two nuclei in the ZF simulation.
ALC results instead shows the largest deviation at low applied field, which is a direct consequence of the presence of an additional local field at the muon site leading to a faster depolarization. Other deviations from the reference are smaller and mostly due to shifts in the resonance field values arising from the combined presence of local fields at the muon and V nuclei. This rules out static local fields at the muon site as the possible origin of the increasing resonance area below $T^*$ in \avs systems.
As a result, also for \kvs, we can conclude that the observed enhancement in the resonance areas is not of magnetic origin and is most likely connected to charge rearrangement that may indirectly modify the dipolar coupling between the muon and the V atoms. 

Fig.~\ref{fig:summary} summarizes the transitions observed with (ALC) \musr in this work and Ref.~\cite{bonfa2024unveiling} and compares them with the anomalies identified by multiple independent experimental investigations, performed with other experimental techniques, between $T^*$ and $T_{CDW}$ for all members of \avs ($A$= Cs, K and Rb) family.
These results have been associated with various physical mechanisms including further modulation of the charge order, loss of rotational symmetry and nematicity, chiral charge order, development of orbital currents and presence of small static magnetic moments. 
Despite the absence of clear thermodynamic signatures of additional phase transitions below the onset of charge order in \avs, a growing body of microscopic measurements reveals distinct anomalies within the charge-ordered state. Our results further support a complex temperature evolution, likely involving both electronic and magnetic channels, and point to a rich sequence of microscopic changes beyond the primary charge-order transition. Developing a comprehensive and coherent microscopic picture that reconciles the diverse experimental observations reported across different techniques remains an important open challenge and warrants further experimental and theoretical investigation.

\section{Conclusion}
We performed ZF and ALC-\musr measurements on \cvs and \kvs systems. 
The ZF relaxation rates observed for \kvs agree well with the previously published results and feature only a small variation at the CDW transition followed by a notable increase just below $T_{CDW}$.
Meanwhile, new ALC results on \cvs reveal a change of resonance field with temperature pointing to a variation of the charge distribution at the V sites at $T^*$ below the CDW transition. On the contrary, this effect is barely visible in \kvs.
In both cases, as well in \rvs, however, a pronounced change in resonance area, generally connected to a variation of the dipolar coupling between the muon and quadrupolar nuclei \cite{Kreitzman1986}, is observed.

From the combined analysis of computational and experimental results we conclude that the measured ALC response cannot be explained solely by the appearance of additional static magnetic fields at the muon sites, and, consistent with previous studies, our results preclude dynamical magnetism in the time-window probed by \musr. 
For \cvs, ALC measurements support a more stringent conclusion: an electronic mechanism alters the local charge distribution in the V plane, thus providing further evidence for a secondary electronic modulation developing within the CDW phase.

\begin{acknowledgments}
We thank Roberto De Renzi, Giuseppe Allodi, Giovanni Pizzi, Ifeanyi John Onuorah, and Ilija Nikolov for fruitful discussions.
The authors gratefully acknowledge the financial support of Consiglio Nazionale delle Ricerche within CNR-STFC Agreement 2021-2027 (N0065606), concerning collaboration in scientific research RB2510399 (\url{https://data.isis.stfc.ac.uk/doi/INVESTIGATION/128218241/}) and RB2520479 (\url{https://topcat.isis.stfc.ac.uk/doi/STUDY/132547683/}) at the ISIS Neutron and Muon Source (UK) of the Science and Technology Facilities Council (STFC).  
The computational resources were provided by the STFC Scientific Computing's SCARF cluster and by the CINECA ISCRA initiative through IsB31\_AER, CNHPC\_1570115 and  IsB31\_MUQCCO. Z.G. acknowledges support from the Swiss National Science Foundation (SNSF) through SNSF Starting Grant (No. TMSGI2${\_}$211750) and through the NCCR Muoniverse, a National Centre of Competence in Research (grant number 51NF-0${\_}$229254). A.K. acknowledges funding support by PRIN project 202243JHMW. S.D.W. and A.C.S. acknowledge support via the UC Santa Barbara NSF Quantum Foundry funded via the Q-AMASE-i program under award DMR-1906325.
\end{acknowledgments}
\bibliography{ref}

\end{document}

\def\avs {AV$_3$Sb$_5$\xspace}
\def\rvs {RbV$_3$Sb$_5$\xspace}
\def\cvs {CsV$_3$Sb$_5$\xspace}
\def\kvs {KV$_3$Sb$_5$\xspace}
\def\V {$^{51}$V\xspace}
\def\Sb {$^{121}$Sb\xspace}
\def\musr {$\mu$SR\xspace}

\title{Evolving charge order in the CDW state of \avs{} - Supplementary Material}

\author{Anshu Kataria}
\affiliation{
Dipartimento di Scienze Matematiche, Fisiche e Informatiche, Universit\`a di Parma, I-43124 Parma, Italy
}

\author{Francis Pratt}
\affiliation{ISIS Pulsed Neutron and Muon Source, Rutherford Appleton Laboratory, Didcot OX11 0QX, U.K.}

\author{Peter J. Baker}
\affiliation{ISIS Pulsed Neutron and Muon Source, Rutherford Appleton Laboratory, Didcot OX11 0QX, U.K.}

\author{Stephen Cottrell}
\affiliation{ISIS Pulsed Neutron and Muon Source, Rutherford Appleton Laboratory, Didcot OX11 0QX, U.K.} 

\author{Miki Bonacci}
\affiliation{PSI Center for Scientific Computing, Theory and Data, 5232 Villigen PSI, Switzerland}

\author{Andrea Capa Salinas} 
\affiliation{Materials Department, University of California Santa Barbara, Santa Barbara, California 93106, USA}

\author{Stephen D. Wilson}
\affiliation{Materials Department, University of California Santa Barbara,Santa Barbara, California 93106, USA}

\author{Zurab Guguchia}
\affiliation{PSI Center for Neutron and Muon Sciences CNM, 5232 Villigen PSI, Switzerland}

\author{Samuele Sanna}
\affiliation{
Dipartimento di Fisica e Astronomia  ``A. Righi'', Universit\`a di Bologna, I-40127 Bologna, Italy }

\author{\hspace{1mm}Pietro Bonf\`a}
\affiliation{Dipartimento di Fisica, Informatica e Matematica, Universit\`a di Modena e Reggio Emilia, Via Campi 213/a, 41125 Modena, Italy}
\affiliation{CNR-NANO S3—Istituto Nanoscienze, I-41125 Modena, Italy }

\date{\today}
\maketitle

\section{MuSR experimental details} 

Zero-field (ZF) and avoided level crossing (ALC) muon spin relaxation and resonance ($\mu$SR) measurements have been performed on high quality powdered \cvs and \kvs samples at the ISIS facility from temperatures ranging from 5 to 150 K using the EMU spectrometer. To reduce the background contribution from muons stopping outside the sample in ZF measurements, we used a kapton mask whereas in ALC measurements, a silver mask is used to reduce the temperature and field dependence of the background.
In the ALC experiments, the muon asymmetry $A(t)$
was time-integrated (TI) from 3.0 to 31 $\mu$s to obtain the values reported in the main text as a function of applied longitudinal field (LF) at various temperatures. The integral is numerically performed on the experimental data by combining all histograms in the specified time interval and, as a consequence, it implicitly contains the exponentially decaying number of counts per histogram bin as a weighting factor.

The positive, spin 1/2, muon interacts with the neighboring nuclear spins $I$ through the dipolar interaction, while the electric field gradient (EFG) at the nuclear sites splits the energy levels of nuclei having spin $I > 1/2$ via the quadrupolar interaction. 
Under applied LF, when the Zeeman splitting of the muon energy levels matches the quadrupolar splitting of one or more nuclei, a resonance dip is expected in the TI spectra (more details below). 
If instead the relaxation rate of LF asymmetry is considered, a peak will be observed at the resonance in the spectra (see next section).

As discussed in the main text, ALC-\musr measurements can be analyzed using two complementary approaches: TI asymmetry and fits to the LF relaxation, each having its own advantages and disadvantages. The analysis based on the TI asymmetry allows a direct comparison with the DFT simulation, but it is more sensitive to extrinsic experiential details such as changes of the total experimental asymmetry due to striction effects of the sample holder, field dependence of other materials explored by the muons ($i.e.,$ muons stopping outside the sample) or muon beam instability. 
The second approach, based on the analysis of LF relaxation rate, is more resilient to variations of the position of the muon beam spot on the sample, but the choice of a functional form to fit the spectra involves several approximations, as discussed in Ref.~\cite{Kreitzman1986}. Here, we use the simplest approach, i.e. a stretched exponential fit of LF data that works well across the resonances but shows clear deviations at low fields, as displayed in Fig.~\ref{rlx_alc_kvs}(a). This complicates the comparison with the simulation.

Finally, we point out that, in general, due to the presence of multiple nuclei with different orientations, the experimentally observed resonance consists of a weighted average of numerous different contributions from various (avoided) level crossings that create a non-trivial shape and width of the resonance dips/peaks. A pedagogical example is given in Ref.~\cite{Bentley_2023}.

\section{ALC relaxation analysis}

\begin{figure*}
\includegraphics[width=1.0\textwidth]{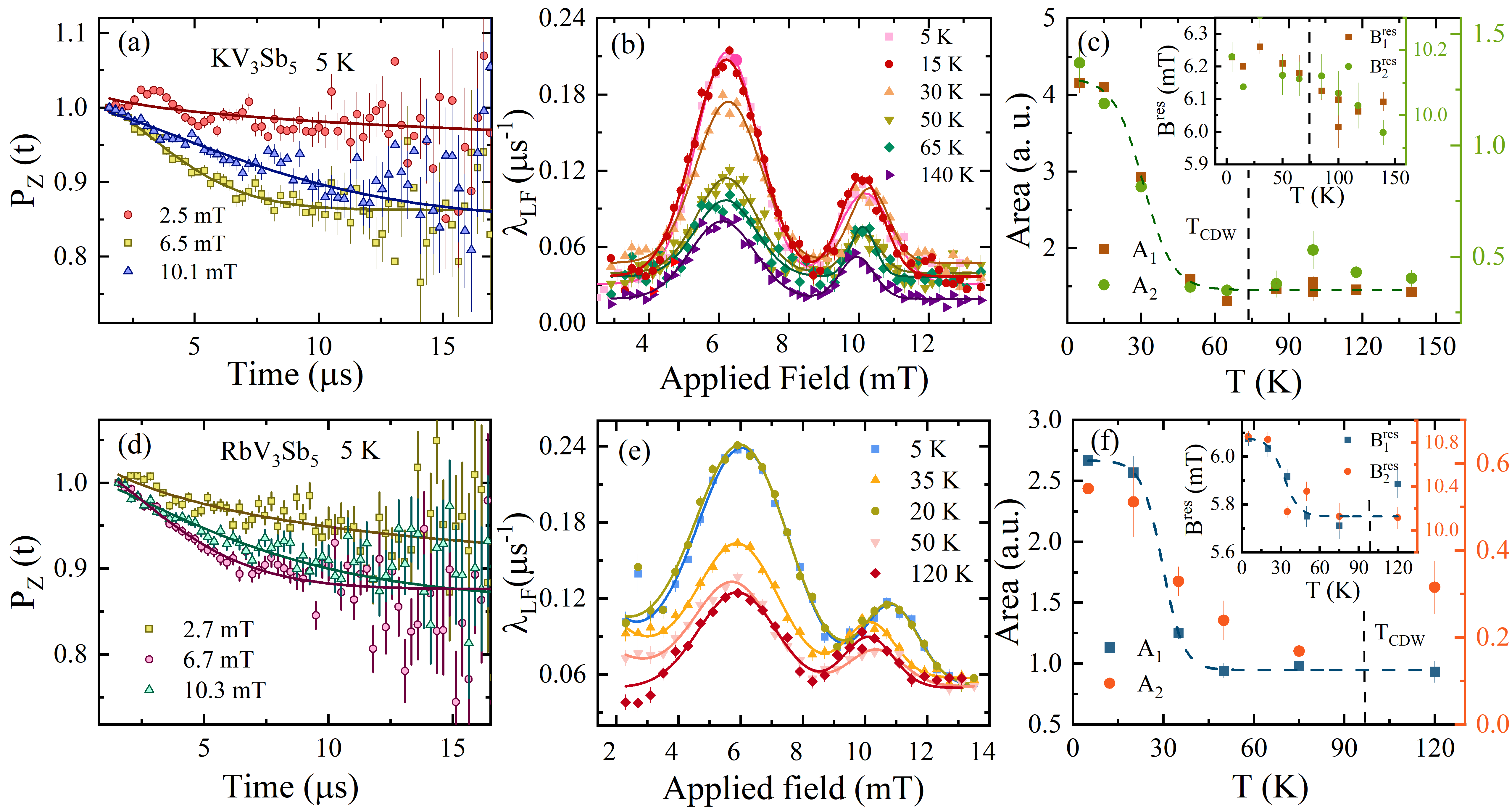}
\caption{The top panels show the analysis based on the LF relaxation rate for \kvs, whereas the bottom panel presents the same analysis for \rvs. Panel (a) and (d) show the time-dependent asymmetry at 5 K under increasing longitudinal field, with solid-colored lines representing the corresponding fit using the Eq. \ref{eq:lf_fit}. Panels (b) and (e) show the fitted $\lambda_{\mathrm{LF}}$ as a function of applied field at different measured temperatures. Here, the solid-colored line represents the fitted Gaussian functions. Finally, the corresponding peak areas obtained from the Gaussian fits are shown in (c) and (f), with the temperature dependence of the resonance field displayed in the respective insets. The dashed lines in both panels serve as guides to the eye, while the vertical black dashed line indicates the CDW transition temperature, $T_{\text{CDW}}$.
\label{rlx_alc_kvs}}
\end{figure*}
In this section we describe the analysis based on the fit of relaxation rates of LF ALC-$\mu$SR, presented in the previous section and mentioned in the main text. Since it provides equivalent information with respect to the TI method, already detailed in the main text, we only consider \kvs, where the temperature dependence of the resonance field is minimal and therefore the use of multiple approaches may be more informative. For comparison, we have also reanalyzed the previous ALC measurements of \rvs polycrystalline sample, published in Ref \cite{bonfa2024unveiling}.
For each LF measurement, the asymmetry spectra are fitted with a stretched exponential function of the form
\begin{equation}
    A_{\text{LF}}(t) = A_0    \exp{(-\lambda_{\text{LF}}t)^{\beta}} + A_{bg}
    \label{eq:lf_fit}
\end{equation}

where $\lambda_{\text{LF}}$ is the relaxation rate, $\beta$ is the exponent of the stretched exponential and $A_0$ and $ A_{bg}$ account for the relaxing and background contributions, respectively. $A_{bg}$ is fixed to the value obtained at resonance at the lowest measured temperature.

The fit of the asymmetry obtained at 5~K for different longitudinal field is shown by a solid-colored line in \ref{rlx_alc_kvs}(a) and (d), for \kvs and \rvs, respectively. For clarity, only the time-spectra with the most pronounced change in relaxation rate are shown.
Finally, the fitted $\lambda_{\text{LF}}$ variation with applied field at different temperature is shown in the Fig. \ref{rlx_alc_kvs}(b) and (e). The observed peak in the spectra are the signature of the resonance with center at the resonance fields, while a similar variation is noted for the correlated  $\beta$ parameter (not shown here).  %
 
We have further fitted the relaxation rate curves shown in Fig.~\ref{rlx_alc_kvs}(b,e) by considering the sum of two Gaussian functions each describing one resonance with a background term. For \kvs, the background is almost flat whereas for \rvs, we considered a field dependent power-law term. 
The resulting best-fit curves are shown by the solid colored lines. The fitted parameters, $i.e.,$ area, and peak position (resonance fields) for the two peaks with their temperature variation is shown in Fig.~\ref{rlx_alc_kvs}(c) and (f), for \kvs and \rvs, respectively. These results are in-align with the ones obtained from the TI asymmetry analysis (Fig 2. of the main text). Notably, for \kvs, the present approach provides a cleaner indication that resonances become temperature independent at very low T and above the CDW transition \footnote{The high-temperature data points (120 and 140 K) are not included in the main text, as these measurements were carried out during a second experimental run. A clear deviation in the TI asymmetry values was observed between the two runs, which arises from variations in the experimental setup. To avoid confusion, only data up to 100 K are presented in the main text.}.
The small change in the ALC resonance field also appears to be less noisy for \kvs, but it remains in quantitative agreement with the results presented in the main text.
Similar to the other compounds of the family, \rvs also exhibits an increasing relaxation rate (or equivalently, the dip in the TI asymmetry)  as the temperature is decreased. In contrast to \kvs, a varying resonance field with a clear temperature onset is observed in \rvs, as shown in the inset of Fig. \ref{rlx_alc_kvs} (f).

\section{First Principles and Spin Hamiltonian simulation details} \label{sec:sims}

The DFT+$\mu$ approach has been used to calculate the muon stopping site in both \cvs and \kvs compounds \cite{bookroberto,blundell2023dft+}. The $\pi$-shifted trihexagonal crystal structure with $Fmmm$ symmetry \cite{PhysRevB.104.195132,kagome.first,PhysRevLett.127.046401} was used to perform all the calculations using the QuantumESPRESSO code \cite{QE-2009, qe_2017,qe2023}. The stable muon sites and the structural perturbation were estimated using the same approach described for \rvs in Ref.~\cite{graham2024depth}, with the exception of a few  details provided below.
PAW pseudopotentials from the PSLibrary \cite{DALCORSO2014337} were used with a plane wave expansion of 80 (800) Ry for plane waves (charge density). The supercells contained 289 atoms, with lattice parameters taken from \cite{PhysRevMaterials.6.015001} and reciprocal space sampling performed on a $3 \times 3 \times 3$ symmetry reduced grid. All structural relaxation calculations were performed using the optB88-vdW exchange-correlation functional to accurately account for the van der Waals interaction.
Supported by previous investigation on these materials \cite{graham2024depth,bonfa2024unveiling,Onuorah2025},
in this work, we only considered the already known best candidate muon sites as a starting point for the structural relaxation.
Finally, although previous analysis on \rvs shows that anharmonicity slightly changes the expectation value for the muon position in $site$ 1 \cite{graham2024depth}, we neglect this effect here, since it is much smaller than the variations observed in both ZF and ALC-\musr measurements as a function of temperature.

Finally, the electric field gradient (EFG) at the various nuclear sites has been calculated with PAW pseudopotentials using the GIPAW code in Quantum Espresso \cite{gipaw}. The numerical solution with EFG of muon depolarization and ALC spectrum is obtained using the Celio approach \cite{Celio_1986}, as implemented in the UNDI code~\cite{undi}.
 
\section{Muon site information}

In the $\pi$-shifted tri-hexagonal phase, three symmetry inequivalent muon sites with similar total energies are identified for both \cvs and \kvs compounds. Their details are summarized in the Table~\ref{tab:muon}, where they are grouped into the set of sites labeled ``site~1''. 
For all of them, the nearest neighbor atom is an in-plane Sb atom, then a Cs/K atom, a hexagon of out-of-kagome-plane Sb atoms and the farthest hexagon of six V atoms in the kagome plane. The results obtained for \cvs and \kvs are in close alignment with the reports of its sister compound \rvs \cite{graham2024depth}. A second set of muon sites (labeled site~2), show a nearest neighbor arrangement that is significantly different as describe in Table~\ref{tab:muon}.
Both sets of sites are depicted in Fig.~\ref{fig:final} (a,b).

\begin{figure}
\includegraphics[width=0.50\textwidth]{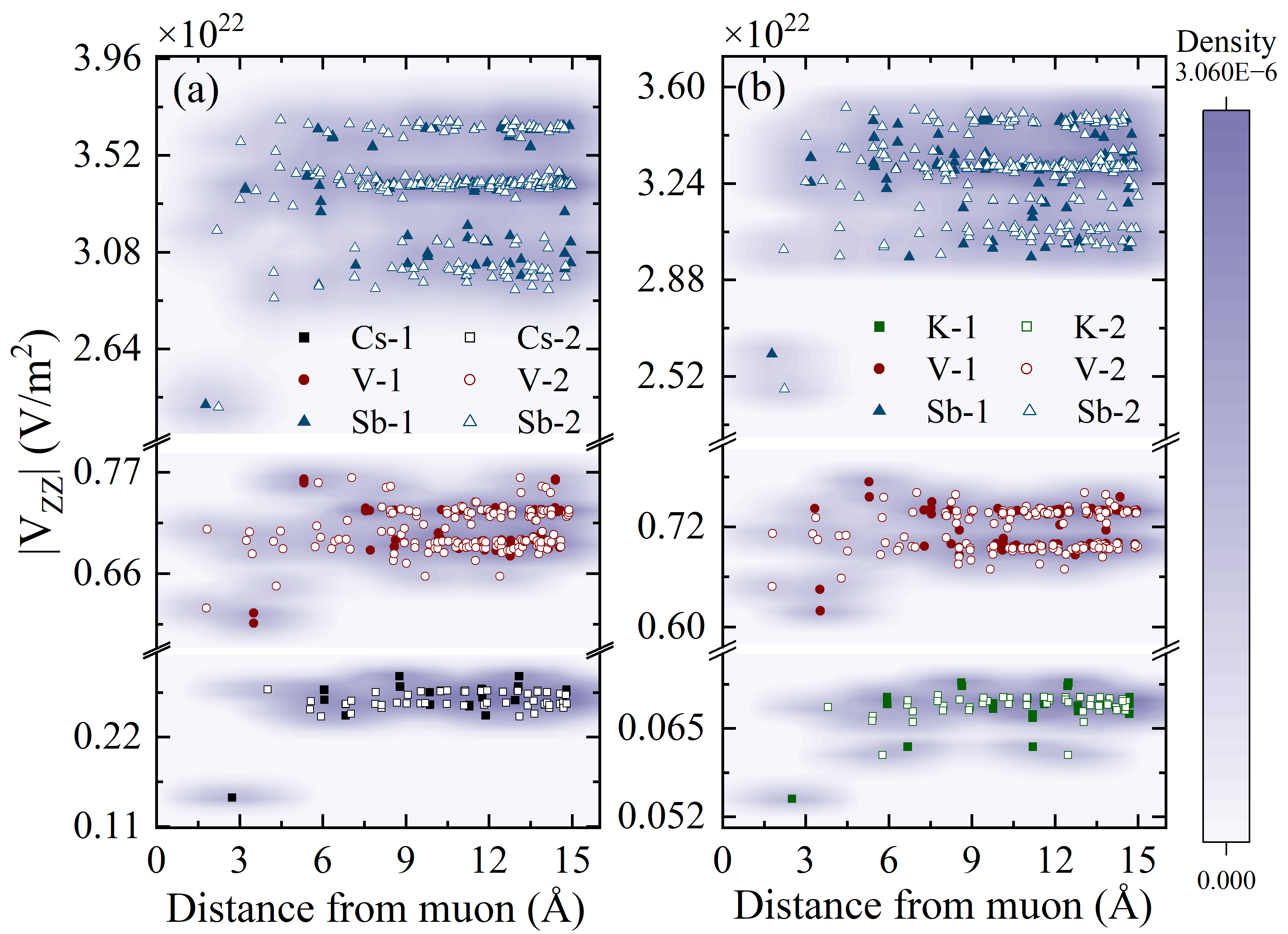}
\caption{Variation of the EFG at different nuclear sites as a function of the distance from the muon for (a) \cvs and (b) \kvs. Solid and hollow symbols represent the EFG values of the nuclei subject to the perturbation of a muon in sites 1 and 2, respectively. The shaded region indicates the density of the corresponding EFG values. }
\label{efg_dis}
\end{figure}

The EFGs at different nuclear sites are plotted as a function of their distance from the implanted muon for sites 1 and 2 and for both compounds in  Fig. \ref{efg_dis}. It is noted that the nearest Sb and Cs/K are strongly perturbed by the muon presence while for the nuclei far from the muon (\textit{i.e.,} the V nuclei), the EFG tensors are only slightly perturbed. Further, the computational prediction for the nuclei far from the muon is close to the experimental values \cite{PhysRevB.107.184106,luo2022possible} probed with NMR.

\begin{table*}[]
    \centering
    \begin{tabular}{c|c|c|c|c|c}\hline

     Compound & Label (hex. ) & Wyckoff pos. & Position (frac. coord) & Energy difference (meV) & Distance from NN ({\AA})/NNN ({\AA})/..\\
           & Wyck. pos. &  & & & \\\hline
     \multirow{10}{*}{\cvs{}} & \multirow{3}{*}{Site 1 (2e)} &  8i& (0,0,0.385) &0 & Sb(1.76) Cs(2.76) 6Sb(3.18-3.20) 6V(3.53-3.56) \\
     & & 16j &(0.125, 0.250, 0.114)& 4 & Sb(1.76) Cs(2.73) 6Sb(3.19) 6V(3.52-3.55) \\
     &  & 8i & (0,0,0.116)& 167 & Sb(1.76) Cs(2.72) 6Sb(3.20) 6V(3.52) \\

      \cline{2-6}
\hspace{0.5cm}
     & \multirow{7}{*}{Site 2 (12o)} & 32p & (0.058, 0.205, 0.055) & 437 & 2V(1.79,1.81) 4Sb(2.18,2.23,3.01,3.03) 3V(3.24-3.45) \\
     & & 32p & & &  \\
     & & 32p & & &  \\
     & & 32p & & &  \\
      & & 32p & & &  \\
     & & 16n & & &  \\
     & & 16n & & &  \\
    
      \hline
      \multirow{10}{*}{\kvs}  & \multirow{3}{*}{Site 1 (2e)} &  8i& (0,0,0.377)&0& Sb(1.78) K(2.52) 6Sb(3.18-3.19) 6V(3.54-3.56)\\
      & &  16j&(0.125, 0.250, 0.123)& 1& Sb(1.78) K(2.51) 6Sb(3.18-3.19) 6V(3.54-3.56)\\
      & &  8i& (0,0,0.123)& 109& Sb(1.78) K(2.50) 6Sb(3.18) 6V(3.52)\\
\cline{2-6}
     & \multirow{7}{*}{Site 2 (12o)} & 32p & ( 0.058, 0.205, 0.058) & 433&  2V(1.79,1.81) 4Sb(2.20,2.23,2.99,3.00) 3V(3.24-3.45) \\
     & & 32p &  &  &   \\
     & & 32p & & &  \\
     & & 32p & & &  \\
      & & 32p & & &  \\
     & & 16n & & &  \\
     & & 16n & & &  \\
    \end{tabular}
    \caption{Lowest energy candidate muon sites found in \cvs and \kvs along with their distance from nearest neighbor (NN) and next nearest neighbor (NNN). Three lowest energy sites form the set of candidates labeled ``site 1'' in the main text. A second set of sites is labeled ``site 2'' using the same criterion of similarity of atomic neighborhood discussed in the main text. Only one symmetry inequivalent site in the set labeled site~2 has actually been computed and appears in the table. Columns contain (from left to right): sample chemical formula, Wyckoff position of the site in the high-temperature Hexagonal structure, Wyckoff position in the low temperature orthorhombic \textit{Fmmm} structure, representative fractional coordinate of muon sites in $Fmmm$ structure used in the simulation, energy difference with respect to the lowest energy one, neighbors of the muon and their distance from it.}
    \label{tab:muon}
\end{table*}

\section{Simulation of ZF polarization functions and comparison with experimental data}

ZF-\musr measurements on \kvs were collected as detailed above, while \cvs results have been previously reported in Ref.~\cite{cvs_zfmuon} and the corresponding data published in the ISIS data repository \cite{isisdatbase}. 
Here, we compare zero-field polarization functions, $P_{\text{ZF}}$(t), for both compounds with \textit{ab initio} predictions including both dipolar interactions between the muon and the neighboring nuclei and quadrupolar interactions at nuclear sites (see Eq.~3 in main text).
 
For these simulations, the dimension of the Hilbert space increases exponentially with the number of nuclei included in the calculation. To obtain a computationally treatable problem without losing quantitative accuracy, only the eight nuclei close to the muon, forming the ASb7 (A=Cs, K) cluster, are considered. This approach has been thoroughly tested in Ref.~\cite{graham2024depth} for \rvs, where it was found that the additional inclusion of the six neighboring V nuclei has only a negligible effect and can therefore be neglected.
Following the approach of Ref.~\cite{graham2024depth}, we adopt an approximate isotope averaging to the (Cs/K)Sb7 cluster. Here, all the nuclei are assumed to be of the same isotope and an average is calculated based on the natural abundances of each isotope. As a result, simulations are performed for $^{121}$Sb,  $^{123}$Sb and $^{133}$Cs, $^{39}$K and $^{41}$K.
Similarly, for the simulation of ALC-\musr resonances, we use instead the cluster ASbV6 for site 1 and Sb2V5 for site 2.

\begin{figure}
\includegraphics[width=0.45\textwidth]{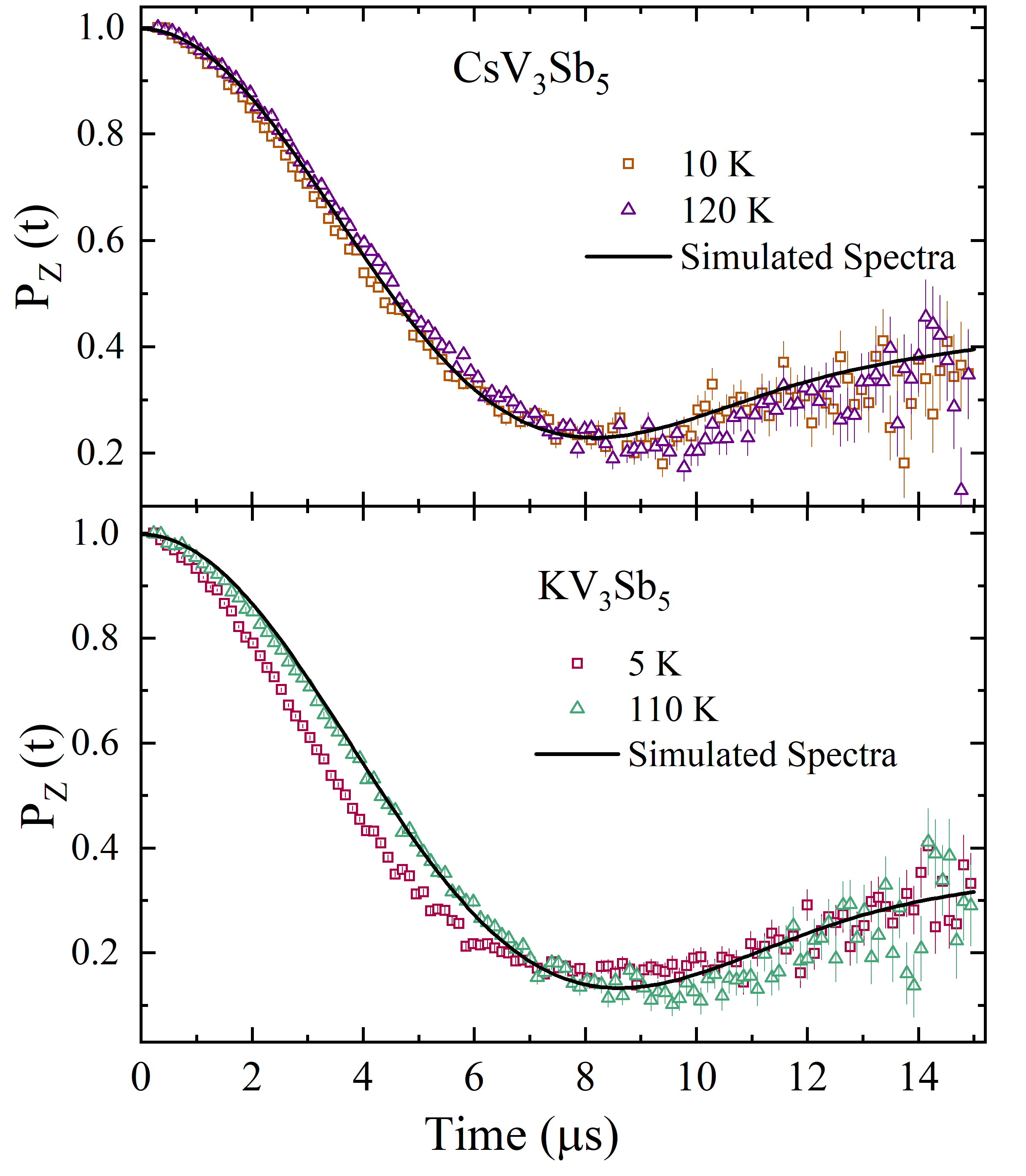}
\caption{Zero-field $\mu$SR time spectra of powdered \cvs and \kvs. The black solid curves for both system represent the $ab$-$initio$ predicted muon asymmetry for muon site 1.}
\label{zf}
\end{figure}

The simulated ZF depolarization spectra for muon site 1 in both \cvs and \kvs are shown in  Fig. \ref{zf}. For \cvs, the predicted spectrum lies between the low and high-temperature curves whereas for \kvs, it shows very good agreement with the experimental data at high-temperature (110 K).
Notice that the ZF asymmetry is almost unaffected by the CDW transition (see main text) and therefore our simulations, obtained considering the low temperature $\pi$-shifted tri-hexagonal structure, match equally well with all measurements collected above $T^*$ for both \cvs and \kvs compounds.

\section{Quadrupolar level crossing resonance in $\mu$SR measurements}

\begin{figure*}
    \centering
    \includegraphics[width=0.9\linewidth]{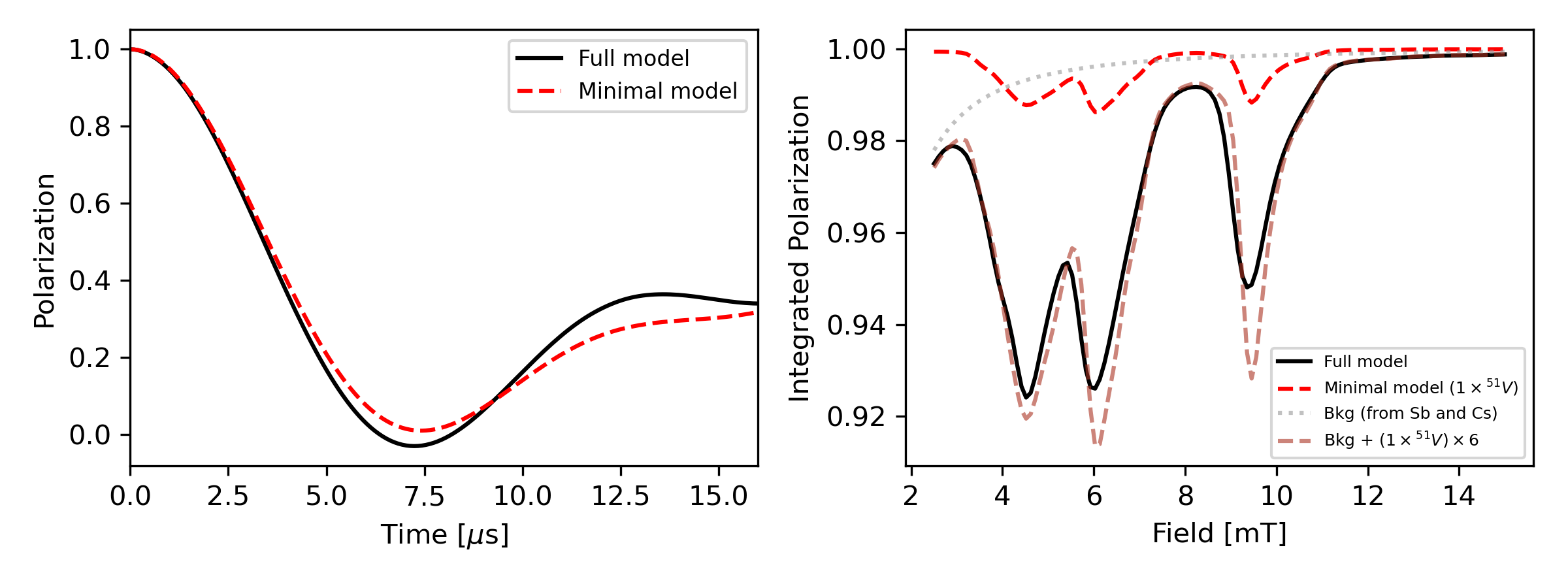}
    \caption{Comparison of the results obtained with the CsSb7 cluster or the SbV6 cluster described in section \ref{sec:sims} and minimal model for \cvs (see text).}
    \label{fig:minimal_model_comparison}
\end{figure*}
In this section we provide some details on the physics of ALC-\musr measurements with a focus on quadrupolar level crossing resonances presented in this manuscript.

\subsection{Resonance field}

Naively, a resonance in the quadrupolar ALC experiments is observed when the Larmor frequency of the muon matches the quadrupolar splitting of the nuclear energy levels. More precisely, an avoided level crossing resonance between a muon and a single quadrupolar nucleus happens when \cite{Kreitzman1986}
\begin{equation}
    \omega_Z ^\mu \pm \omega_q ^i \pm \omega_z ^i=0
\end{equation}
where $\omega_Z ^\mu = \gamma_\mu B^\mu _\text{loc}$ is the Zeeman frequency of the muon,
$\omega_q^i$ and $\omega_z ^i=\gamma_i B^{(i)} _\text{loc}$ are the quadrupolar and Zeeman frequencies of the $i$-th nucleus, respectively.
The Zeeman terms may be affected by local fields, while the quadrupolar term changes due to a variation of the EFG tensor at the nucleus.
Notice that, for $^{51}$V nuclei, the ratio $\gamma_\mu / \gamma_V \sim 12$, so the role of the Zeeman contribution from the nuclei is limited, thus leading to the resonance condition $\omega_Z ^\mu \simeq \pm \omega_q ^i$, described at the beginning of this section.
In an \musr experiment, $\omega_Z ^\mu$ is tuned by applying an external field until it matches $\omega_q ^i$. The muon is sensitive to local fields, therefore the applied external field may add up with an internal field of electronic origin, if present. The magnitude of such an internal field can be easily quantified from the ZF measurements, as discussed in the main text.
By combining ZF and ALC results, we immediately understand that in \cvs, the shift in resonance field at $T^*$ cannot be simply due to the presence of an additional magnetic field, since the increased relaxation in ZF measurement would corresponds to an increase of $\leq$~0.1 mT \cite{cvs_zfmuon,PhysRevResearch.4.023244}, which is more than 10 times smaller than the shift observed in the resonance field ($\sim$2mT).

A second possibility to account for the resonance field shift is that  $\omega_z ^i$ for V nuclei changes with temperature, while all other terms remain constant (in particular, the local field distribution at the muon site). This scenario would require a local field variation of at least 10 mT at the V nuclei, which corresponds to a change of $\approx$100 kHz in an NMR experiment. Such a change is well within the NMR resolution, however, only a small change in the $^{51}V$ resonance width has been reported without any measurable frequency shift around $T^*$ \cite{PhysRevB.107.184106,luo2022possible}, thus discarding the above case.

\subsection{Area of the dip}

In the simplified picture of one muon and one nucleus discussed extensively by Kiertzman \cite{Kreitzman1986}, the area of the resonance dip (or equivalently the relaxation rate) is directly linked to two quantities: the dipolar coupling between the muon and the nucleus contributing in the resonance ($^{51}$V in the present case) and the relative alignment of the quadrupolar and dipolar tensors.

Both quantities affect the strength of dipolar terms responsible for the cross-relaxation of the muon spin polarization and can vary either due to the change in the muon-nuclei geometry for the dipolar contribution or due to charge re-distribution for the quadrupolar contribution.

In the present case, however, nuclear movements are found to be limited by previous high resolution x-ray scattering measurements \cite{PhysRevB.105.195136, Scagnoli_2024}, and also by the small change observed in ZF measurements. Under static conditions ($i.e.,$ assuming that only site 1 is occupied and muon diffusion is negligible) and in the absence of electronic contributions, the small temperature dependence of the ZF asymmetry implies that the dipolar interaction between the muon and the nearest Sb nucleus must remain almost constant.
Within this assumptions, the ZF measurements are therefore ``constraining'' the muon to stay roughly in the same position at all temperatures. A possible explanation of the phenomenon must therefore release some of the previous assumptions, possibly allowing the EFG at nuclear sites to change or the presence of dynamical effect involving the muon.

\subsection{Accuracy of the minimal model}

\begin{figure*}
    \centering
    \includegraphics[width=0.99\linewidth]{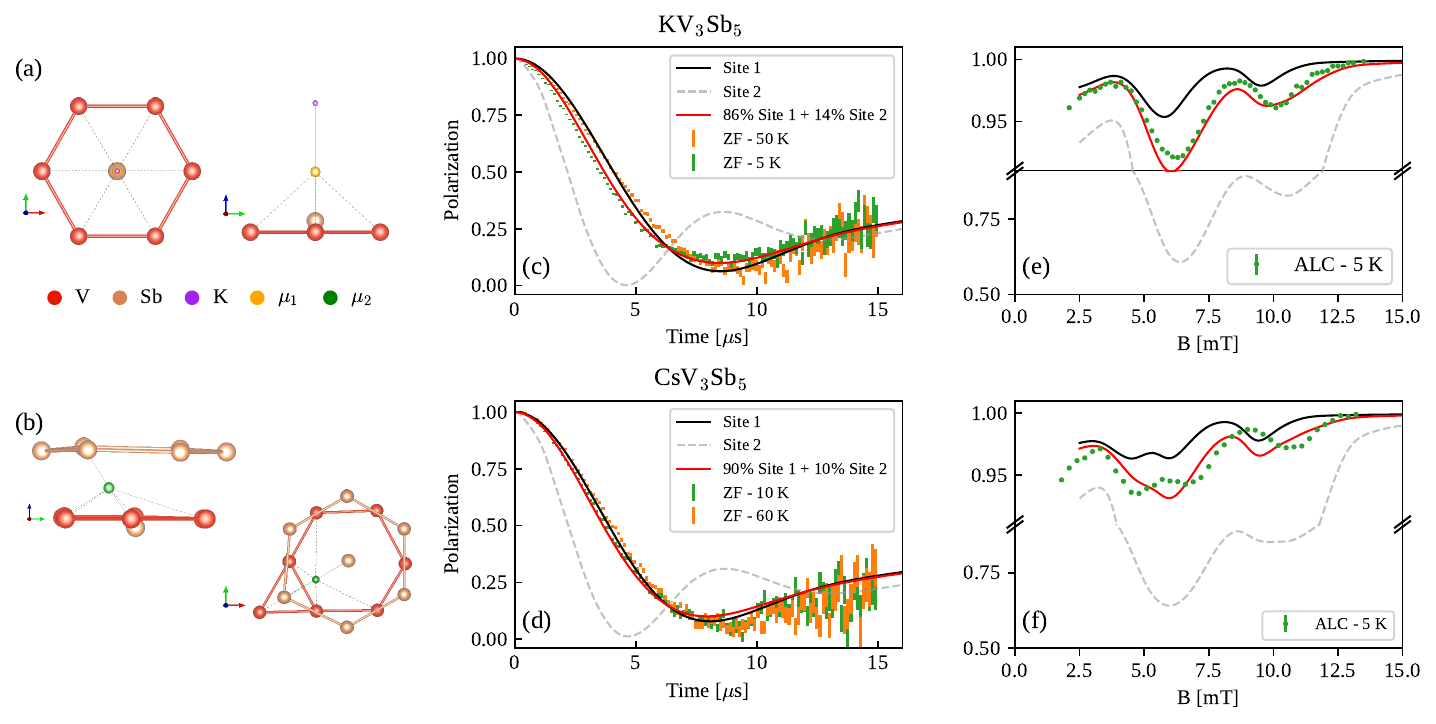}
    \caption{(a) Muon site 1 is shown in two different orientations. (b) Muon site 2, also shown in same two orientations. (c) ZF polarization function for the two sites in \kvs. Experimental data above and below $T^*$ are also shown for comparison. A baseline component has been subtracted and the remaining asymmetry is normalized to 1. The red line shows a weighted average of the polarization function for site 1 and 2. The weights, reported in the legend, have been manually selected to roughly match the variation observed at low-T for both ZF and ALC measurements. (d) same for \cvs. (e) Simulated ALC resonances in \kvs for both site 1 and site 2 and weighted average, in red, with the same weights used for the ZF panel. The experimental data at 5~K are shown for comparison. (f) same for \cvs.}
    \label{fig:final}
\end{figure*}

In order to obtain a tractable Hamiltonian for the description of the effect of local fields at the muon and V sites, in the main text we introduce a minimal model for the description of both the ZF and the ALC-\musr measurements.
For the minimal model, in zero field we consider only the two nearest neighbors nuclei of the muon, one Sb and Cs/K, while for the ALC, we only consider one V nucleus.
The full model (our reference in this analysis) instead considers the interaction between the muon, 7 Sb nuclei and one $A$=K,Rb,Cs for ZF relaxation, while for the simulation of ALC spectra, the 6 second nearest neighbors Sb nuclei of the muon can be safely neglected, thus leaving 8 nuclei in total ($^{121}$Sb or $^{123}$Sb, $6 \times ^{51}$V and $^{133}$C or $^{39/41}$K or $^{85/87}$Rb).
For the purpure of comparing the minimal model with the full model, we skipped isotope averages and considered only the most abundant isotopes for each nucleus.
The comparison between the two models is shown in Fig.~\ref{fig:minimal_model_comparison} with the results obtained for \cvs. A good agreement is obtained for ZF measurements for times smaller than 6 $\mu s$. The trend deviates slightly at higher times due to the limited size of the Hilbert space which constraints the decoherence process.
Most importantly, the ALC resonance is also well reproduced by simply scaling the effect obtained for a single V nucleus by a factor 6 and adding it on top of the "background" produced by the other two nuclei (Sb and Cs).

\section{Towards a quantitative description of ALC experiments}

A complete, microscopic and quantitative understanding of the origin of the temperature dependence of the ALC experiments is beyond the scope of the present work.
Still we may attempt to frame a few key points that could be relevant for future analysis on these systems.

A key quantity governing the relaxation rate in the present ALC experiment is the dipolar interaction between the muon and V nuclei.
Previous DFT+$\mu$ simulations \cite{graham2024depth, Onuorah2025}, identify multiple candidate muon sites, with the second best set of candidates
\footnote{As already mentioned for site 1, also for site 2 the symmetry equivalent positions with Wyckoff site 12o in the hexagonal structure split up into 7 symmetry inequivalent sites with 32p or 16m Wyckoff position and roughly the same surroundings. To simplify the analysis and the description, we only consider one of them here.} also being located close to the kagome plane, as shown in Fig.~\ref{fig:final}(b), with a total energy around 400~meV higher than the lowest energy site (Table~\ref{tab:muon}).

For both  \kvs and \cvs compounds, the second muon-site gives a faster relaxation rate in ZF measurements as there are now two nearest neighboring Sb atoms about $\sim$ 2.2 \AA ~away from the moun (Fig.~\ref{fig:final}(c) and (d)).
At the same time, the muon becomes closer to two V atoms. Both these nuclei contribute to the ZF relaxation, but, more importantly, their dipolar with the muon increases by a factor $\sim 8$ with respect to the other site (the distance, $\sim 1.8$~\AA, is roughly halved), thus justifying the stronger resonance dip.

Surprisingly, according to our simulations, even when the distance between the muon and V atoms is substantially reduced, the perturbation of the EFG tensors at V atoms remains small (see Fig.~\ref{efg_dis}). This is also indirectly but quantitatively shown in Fig.~\ref{fig:final} by the comparison of the resonance field distributions (continuous black and dashed gray lines) at the two site.

Allowing a small muon occupation at site 2 at low temperature provides a consistent description of both ZF and ALC measurements for \kvs, as shown in Fig.~\ref{fig:final}(c) and (e).
The agreement is only qualitative for \cvs, for which the contribution from the second site leads to redistribution of the intensity of the first two dips that is not observed in the experiment.

We stress that this preliminary analysis is solely intended to provide a plausible set of contributions for the descriptio of the experimental observations. 
It should in fact be noted that the static description 
discussed here would also fail to explain the departure from the Gaussian relaxation visible at early times in ZF measurements.
The presence of more than one site is however qualitatively compatible with previous measurements performed on \cvs{} single crystals \cite{Gautreau} indicating that the relaxation rate becomes progressively more isotropic
at low-T, roughly starting from what we identify here as $T^*$.

Finally and most importantly, we also point out that an intermediate temperature transition has also been observed in many other experiments such as NMR, STM, ARPES, Raman spectroscopy and others, where a charge redistribution transition and the emergence of local moments is suggested inside the CDW state \cite{luo2022possible,PhysRevX.14.031015,zhao2021cascade,Gui2025}.

It is plausible that the response observed by \musr is a consequence of an intrinsic transition that perturbs the delicate balance of strength and symmetry of the (mostly Coulombic) interactions shaping the muon wavefunction across the various phase transitions characterizing this material.

\bibliography{ref}

%% file: tikzfig.tex
\definecolor{left} {HTML}{005566}
\definecolor{right} {HTML}{992200}

\definecolor{magnet} {HTML}{000000}
\definecolor{charge} {HTML}{000000}

\begin{tikzpicture}

    \node [shading = axis,rectangle, left color=left, right color=left!30!right, anchor=north, minimum width=11cm, minimum height=0.5cm] at (5.5,6) {} ;

    \node [shading = axis,rectangle, left color=left, right color=left!30!right, anchor=north, minimum width=11cm, minimum height=0.5cm] at (5.5,4) {} ;

    \node [shading = axis,rectangle, left color=left, right color=left!30!right, anchor=north, minimum width=11cm, minimum height=0.5cm] at (5.5,2) {} ;
   \draw[thick, ->] (0, 6.2) -- (11.1, 6.2) node[right] {Temperature (K)};
\foreach \x/\label in {0/0,2/20,4/40,6/60,8/80,10/100}
{
    \draw[thick] (\x,6.2) -- (\x,6.4);
    \node[below] at (\x,6.8) {\label};
}
  \node at (-1, 1.75) [align=center, font=\small] {KV$_3$Sb$_5$};
  \node at (-1, 3.75) [align=center, font=\small] {RbV$_3$Sb$_5$};
  \node at (-1, 5.75) [align=center, font=\small] {CsV$_3$Sb$_5$};

 \node at (80mm, 1) [font=\small] {$T_{CDW} \sim 78$ K};
 \draw[white,line width=2mm, opacity=0.5]   (78mm, 1.5) -- (78mm, 2);

 \draw[orange,line width=4mm, opacity=0.5]   (65mm, 1.75) -- (15mm, 1.75);

 \node at (60mm, 1.) [font=\small, align=center] {\cite{acs.nanolett.4c01050,jiang2023observation}};
 \draw[->, magnet, line width=0.25mm, opacity=0.9]   (60mm, 1.3) -- (60mm, 1.5);

  \node at (2mm, 1) [font=\small] {$T_{SC} \sim 1$ K};
 \draw[white,line width=2mm, opacity=0.5]   (1mm, 1) -- (1mm, 2);

 \node at (104mm, 3.2) [font=\small] {$T_{CDW} \sim 104$ K};
 \draw[white,line width=2mm, opacity=0.5]   (104mm, 3) -- (104mm, 4);

 \draw[orange,line width=4mm, opacity=0.5]   (50mm, 3.75) -- (20mm, 3.75);

 \node at (35mm, 3.) [font=\small, align=center] {\cite{Wang_2023}};
 \draw[->, magnet, line width=0.25mm, opacity=0.9]   (35mm, 3.3) -- (35mm, 3.5);

 \node at (50mm, 3.) [font=\small, align=center] {\cite{PhysRevB.105.245123}};
 \draw[->, magnet, line width=0.25mm, opacity=0.9]   (50mm, 3.3) -- (50mm, 3.5);

  \node at (2mm, 3.2) [font=\small] {$T_{SC} \sim 1$ K};
 \draw[white,line width=2mm, opacity=0.5]   (1mm, 3.4) -- (1mm, 4);

 \node at (94mm, 5.2) [font=\small] {$T_{CDW} \sim 94$ K};
 \draw[white,line width=2mm, opacity=0.5]   (94mm, 5) -- (94mm, 6);

 \draw[orange,line width=4mm, opacity=0.45]   (45mm, 5.75) -- (15mm, 5.75);

 \node at (30mm, 5.) [font=\small, align=center] {\cite{Gui2025}};
 \draw[->, magnet, line width=0.25mm, opacity=0.9]   (30mm, 5.3) -- (30mm, 5.5);

 \node at (20mm, 4.6) [font=\small, align=center] {\cite{PhysRevX.14.031015}};
 \draw[->, magnet, line width=0.25mm, opacity=0.9]   (20mm, 5.0) -- (20mm, 5.5);
 
 \node at (35mm, 4.6) [font=\small, align=center] {\cite{nie2022charge, PhysRevB.105.L201109, li2023unidirectional, Song2022,wei2024three,guo2022switchable}};
 \draw[->, magnet, line width=0.25mm, opacity=0.9]   (35mm, 5) -- (35mm, 5.5);

 \node at (45mm, 5.) [font=\small, align=center] {\cite{feng2026nmrevidenceloopcurrentstate}};
 \draw[->, magnet, line width=0.25mm, opacity=0.9]   (45mm, 5.3) -- (45mm, 5.5);

 \node at (55mm, 4.6) [font=\small, align=center] {\cite{Gui2025, zhao2021cascade,li2023unidirectional}};
 \draw[->, magnet, line width=0.25mm, opacity=0.9]   (50mm, 5.0) -- (50mm, 5.5);

\node at (70mm, 5) [font=\small, align=center] {\cite{PhysRevB.105.155106}};
 \draw[->, magnet, line width=0.25mm, opacity=0.9]   (70mm, 5.3) -- (70mm, 5.5);

  \node at (2mm, 5.2) [font=\small] {$T_{SC} \sim 2.5$ K};
 \draw[white,line width=5mm, opacity=0.5]   (1.25mm, 5.4) -- (1.25mm, 6);

  \end{tikzpicture}